\documentclass[aps,pre,11pt,superscriptaddress,showpacs]{revtex4}
\usepackage{graphicx}
\usepackage{epsfig}
\usepackage{amsmath,amssymb}
\usepackage{color}

\def\eq#1{Eq.~(\ref{#1})}
\def\fig#1{Fig.~\ref{#1}}
\def\sec#1{Sec.~\ref{#1}}

\DeclareMathOperator{\csch}{csch}

\begin{document}

\title{Viscoelastic response of contractile filament bundles}
\author{Achim Besser}
\affiliation{University of Heidelberg, Bioquant, Im Neuenheimer Feld 267, 69120 Heidelberg, Germany}
\affiliation{Harvard Medical School, Department of Cell Biology, 240 Longwood Ave, Boston, MA 02115, USA}
\author{Julien Colombelli}
\affiliation{European Molecular Biology Laboratory, Cell Biology and Biophysics Unit, Meyerhofstrasse 1, 69117 Heidelberg, Germany}
\affiliation{Institute for Research in Biomedicine, Baldiri Reixac 10, 08028 Barcelona, Spain}
\author{Ernst H. K. Stelzer}
\affiliation{European Molecular Biology Laboratory, Cell Biology and Biophysics Unit, Meyerhofstrasse 1, 69117 Heidelberg, Germany}
\affiliation{Goethe University, Frankfurt Institute for Molecular Life Sciences, 60323 Frankfurt am Main, Germany}
\author{Ulrich S. Schwarz}
\email[]{Ulrich.Schwarz@bioquant.uni-heidelberg.de}
\affiliation{University of Heidelberg, Bioquant, Im Neuenheimer Feld 267, 69120 Heidelberg, Germany}
\affiliation{University of Heidelberg, Institute for Theoretical Physics, Philosophenweg 19, 69120 Heidelberg, Germany}

\date{\today}

\begin{abstract}
  The actin cytoskeleton of adherent tissue cells often condenses into
  filament bundles contracted by myosin motors, so-called stress
  fibers, which play a crucial role in the mechanical interaction of
  cells with their environment. Stress fibers are usually attached to
  their environment at the endpoints, but possibly also along their
  whole length. We introduce a theoretical model for such contractile
  filament bundles which combines passive viscoelasticity with active
  contractility.  The model equations are solved analytically for two
  different types of boundary conditions.  A free boundary corresponds
  to stress fiber contraction dynamics after laser surgery and results
  in good agreement with experimental data.  Imposing cyclic varying
  boundary forces allows us to calculate the complex modulus of a
  single stress fiber.
\end{abstract}


\maketitle

\section{Introduction}

The actin cytoskeleton is a dynamic filament system used by cells to
achieve mechanical strength and to generate forces. In response to
biochemical or mechanical signals, it switches rapidly between
different morphologies, including isotropic networks and contractile
filament bundles. The isotropic state of crosslinked passive actin
networks has been studied experimentally in great detail, for example
with microrheology \cite{Gittes97,Gardel04}.  Similar approaches have
been applied to actively contracting actin networks \cite{Mizuno07,koenderink_active_2009}
and live cells \cite{Fabry01,Fernandez06}. However, less attention has
been paid to the mechanical response of the other prominent morphology
of the actin cytoskeleton, namely the contractile actin bundles, which
in mature adhesion appear as so-called stress fibers. During recent
years, it has become clear that stress fibers play a crucial role not
only for cell mechanics, but also for the way adherent tissue cells
sense the mechanical properties of their environment
\cite{Discher05,vogel_local_2006,geiger_environmental_2009}. Thus it
is important to understand how passive viscoelasticity and active
contractility conspire in stress fibers.

\begin{figure}
 \centering \includegraphics[width=0.90\textwidth]{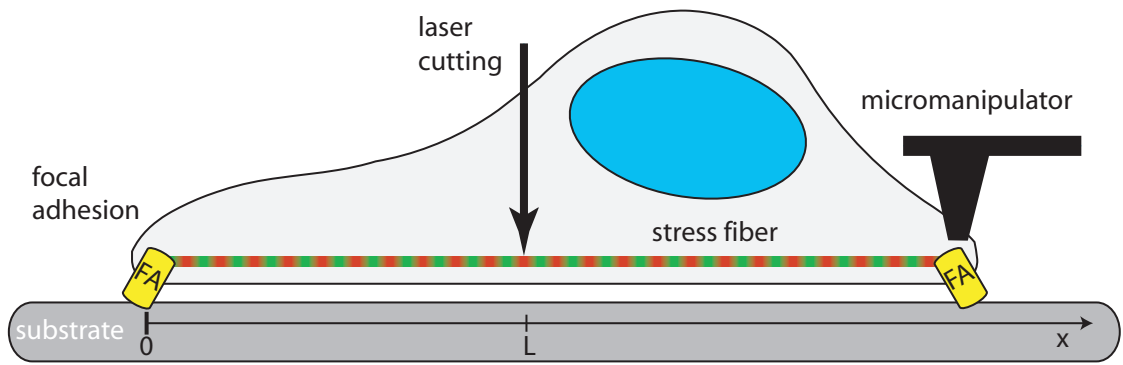}
  \caption{(Color online) Contractile filament bundles are
very prominent in cell adhesion. Stress fibers typically
connect two focal adhesions and through their contraction, the cell
can probe the mechanical properties of the substrate. The
viscoelastic properties of stress fibers can be probed by
laser cutting or by cyclic loading through a micromanipulator.}
\label{overview}
\end{figure}

Stress fibers are often mechanically anchored to sites of cell-matrix
adhesion, are contracted by non-muscle myosin II motors and have a
sarcomeric structure similar to muscle
\cite{pellegrin_actin_2007,Peterson04}, as shown schematically in
\fig{overview}. However, their detailed molecular structure is much
less ordered than in muscle. In particular, stress fibers in live
cells continuously grow out of the focal adhesions
\cite{endlich_movement_2007} and tend to tear themselves apart under
the self-generated stress \cite{smith_zyxin-mediated_2010}. Up to now,
the mechanical response of stress fibers has been measured mainly
isolated from cells
\cite{katoh_isolation_1998,Deguchi06,Matsui09}. Recently, pulsed
lasers have been employed to disrupt single stress fibers in living
cells
\cite{kumar_viscoelastic_2006,uss:colo09,tanner_dissecting_2010}. By
using the intrinsic sarcomeric pattern or an artificial pattern
bleached into the fluorescently labeled stress fibers, the contraction
dynamics of dissected actin stress fibers has been resolved with high
spatial and temporal resolution along their whole length
\cite{uss:colo09}. These experiments showed that dissected stress
fibers contract non-uniformly and that the total contraction length
saturates for long fibers, suggesting that stress fibers in adherent
cells are not only attached at their endpoints, but also along their
whole length. In the same study, cyclic forces have been applied to
stress fibers by an AFM cantilever, mimicking physiological conditions
like in heart, vasculature or gut. \fig{overview} shows schematically
how laser cutting and micromanipulation are applied to an adherent
cell.

Early theory work on stress fibers focused on the dynamics of
self-assembly leading to a stable contractile state
\cite{kruse_actively_2000,kruse_self-organization_2003}.  Later more
detailed mechanical models have been developed and parametrized by
experimental data
\cite{Besser07,Stachowiak08,Luo08,uss:colo09,Stachowiak09,Russell09,uss:bess10a}.
Here we investigate a generic continuum model for the mechanics of
contractile filament bundles and show that it can be solved
analytically for the boundary conditions corresponding to stress fiber
laser nanosurgery and cyclic pulling experiments. Our analytical
results can be easily used for analyzing experimental data. For
relaxation dynamics after laser cutting, our model predicts unexpected
oscillations. We reevaluate data obtained earlier from laser cutting
experiments \cite{uss:colo09} and indeed find evidence for the
predicted oscillations.

This paper is organized as follows. In \sec{sec_model} we introduce
our continuum model, including the central stress fiber equation,
\eq{eq_model_nondim}. The stress fiber equation is a partial
differential equation with mixed spatial and temporal derivatives.  In
order to solve it analytically, in \sec{sec_model_solution} we
discretize this equation in space. This results in a system of
ordinary differential equations, which can be solved in closed form by
an eigenvalue analysis. In \sec{sec_cont_limit}, we take the continuum
limit of this solution, thus arriving at the general solution of the
continuum model.  This general solution is given in
\eq{eq_general_solution}, with the corresponding spectrum of
retardation times given in \eq{eq_retardation_times_cont}.  In
\sec{sec_laser_cutting} and \sec{sec_cyclic_loading}, we specify and
discuss the general solution for the boundary conditions appropriate
for laser cutting and cyclic loading, respectively. In
\sec{sec_discussion}, we close with a discussion.

\section{Model definition and solution}
\label{sec_model} 

We model the effectively one-dimensional stress fiber as a
viscoelastic material which is subject to active myosin contraction
forces and which interacts viscoelasticly with its surrounding. In the
framework of continuum mechanics, the fiber internal viscoelastic stress
is given by the viscoelastic constitutive equation \cite{b:pipkin86}:
\begin{equation}
\label{eq_constitutive_eq}
 \sigma(t)=\int_{-\infty}^t G_{int}(t-t')\dot{\epsilon}(t')dt'\ ,
\end{equation}
where $\sigma=\sigma_{xx}$ and $\epsilon=\epsilon_{xx}=\partial_x u$
denote the relevant components of the stress and strain tensors,
respectively. $u(x,t)$ denotes the displacement along the fiber and
$G_{int}$ is the internal stress relaxation function. In addition to the
viscoelastic stress, the fiber is subject to myosin contractile stress
$\sigma_m$, which we characterize by a linear stress-strain rate
relation ${\sigma_m=\sigma_s(1+\partial_x \dot{u} /
  \dot{\epsilon}_0)}$. $\dot{\epsilon}_0$ denotes the
strain rate of an unloaded fiber and $\sigma_s$ is the maximal
stress that the molecular motors generate under stalling conditions. In addition to the fiber
internal stresses, viscoelastic interactions with the surrounding lead
to body forces, $f_{ext}$, that act over a characteristic length $a$
along the fiber and resist the fiber movement:
\begin{equation}
  f_{ext}=-\frac{1}{a}\int_{-\infty}^t G_{ext}(t-t')\dot{u}(t')dt'\ .
\end{equation}
Here, $G_{ext}$ is the external stress relaxation function.  In the
following we assume that both internal and external stress relaxation functions have the
characteristics of a Kelvin-Voigt material:
\begin{equation}
G(t)=K\theta(t)+2\eta \delta(t)
\end{equation}
where $K$ and $\eta$ are elastic and viscous parameters, respectively,
and $\theta(t)$ and $\delta(t)$ denote the Heaviside step and Dirac
delta function, respectively. The chosen Kelvin-Voigt model is the
simplest model for a viscoelastic solid that can carry load at
constant deformation over a long time. Note that $K_{ext}$ represents
the elastic foundation of the stress fibers revealed by the laser
cutting experiments \cite{uss:colo09}, while $\eta_{ext}$ represents
dissipative interactions between the moving fiber and the cytoplasm.

Our central equation (the \textit{stress fiber equation}) follows from
mechanical equilibrium, $\partial_x \left(\sigma
+\sigma_{m}\right)+f_{ext}=0$, which results in the following partial
differential equation:
\begin{equation} \label{eq_model_nondim}
    \partial_x^2\dot{u}+\partial_x^2 u-\Gamma\dot{u}-\kappa u=0\ .
\end{equation}
This equation has been written in non-dimensional form using the
typical length scale $a$, the time scale
$\tau=\eta_{int}/K_{int}+\sigma_s/(\dot{\epsilon}_0 K_{int})$, the
force scale $f_0=K_{int}$, the non-dimensional ratio of viscosities
$\Gamma=a \eta_{ext}/(\eta_{int}+\sigma_s/\dot{\epsilon}_0)$ and the
non-dimensional ratio of stiffnesses $\kappa=a K_{ext}/K_{int}$.
Eq.~(\ref{eq_model_nondim}) has been derived before via a different
route, namely as the continuum limit of a discrete model representing
the force balance in each sarcomeric element of a discrete model
\cite{Besser07,uss:colo09,uss:bess10a}. However, the pure continuum
viewpoint taken here seems at least equally valid, because stress
fibers are more disordered than muscle and because the interactions
with the environment represented by $K_{ext}$ and $\eta_{ext}$ are
expected to be continuous along the stress fiber. In general, the
stress fiber equation (\ref{eq_model_nondim}) can be solved
numerically with finite element techniques \cite{uss:bess10a}. In this
paper, we show that it also can be solved analytically.

In order to solve the stress fiber equation, we have to impose
boundary and initial conditions. We impose the boundary conditions
that the fiber is firmly attached at its left end at $x=0$, and is
pulled with a certain boundary force $f_b(t)$ at its right end at
$x=L/a=l$:
\begin{equation} \label{eq_model_nondim_BC}
u(0,t) = 0\,\,\,\,\,\,\textrm{and}\,\,\,\,\,\,{\partial_x}\dot{u}(l,t)+{\partial_x}u(l,t)+f_s=f_b(t)\ .
\end{equation}
The boundary condition for $x=l$ describes the balance of forces at
the right end of the fiber where $f(t):=f_s-f_b(t)$ is the difference
between the myosin stall force $f_s=\sigma_s/f_0$ and the externally
applied boundary force $f_b(t)$. As initial condition, we 
simply use $u(x,0)=0$, that is vanishing displacement.

Before we derive the general model solution, we briefly discuss the
special case $\kappa=\Gamma$. This case can be easily solved and
gives first insight into the solution for the displacement field
$u(x,t)$. With the definition $h=\dot{u}+u$, the
partial differential equation \eq{eq_model_nondim} becomes
a homogeneous linear ordinary differential equation:
\begin{equation} \label{eq_ODE_for_h}
 \partial_{x}^2 h-\kappa h=0
\end{equation}
with the boundary conditions $h(0)=0$ and
$\partial_{x} h(l) + f(t)=0$. It can be solved by
an exponential ansatz and thus leads to a 
inhomogeneous linear ordinary differential equation
$\dot{u}+u=h$ for $u$, with the initial condition
$u(x,0)=0$. The final solution reads
\begin{equation} \label{eq_solution_kappa_eq_Gamma} 
\displaystyle u(x,t)=-\frac{\sinh(x\sqrt{\kappa})} {\sqrt{\kappa} \cosh(l\sqrt{\kappa})} \int_0^t f(t')e^{-(t-t')}dt'\ .
\end{equation}
Laser cutting experiments correspond to the situation where the
externally applied boundary forces vanish, that is
$f(t)=f_s=\textrm{const}$. Then the integral
in \eq{eq_solution_kappa_eq_Gamma} is trivial and thus the special case
$\kappa=\Gamma$ leads to a retardation process with a single
retardation time $\tau$. The largest, always negative displacement
given by $-\frac{f_s}{\sqrt{\kappa}}$ occurs at $x=l$, where the fiber
was released by the laser cut. The magnitude of the displacement
decreases exponentially with increasing distance from this point and the
typical length scale of this decay is given by $a/\sqrt{\kappa}$.

\section{Solution of the discretized model}
\label{sec_model_solution} 

\begin{figure}
 \centering \includegraphics[width=0.90\textwidth]{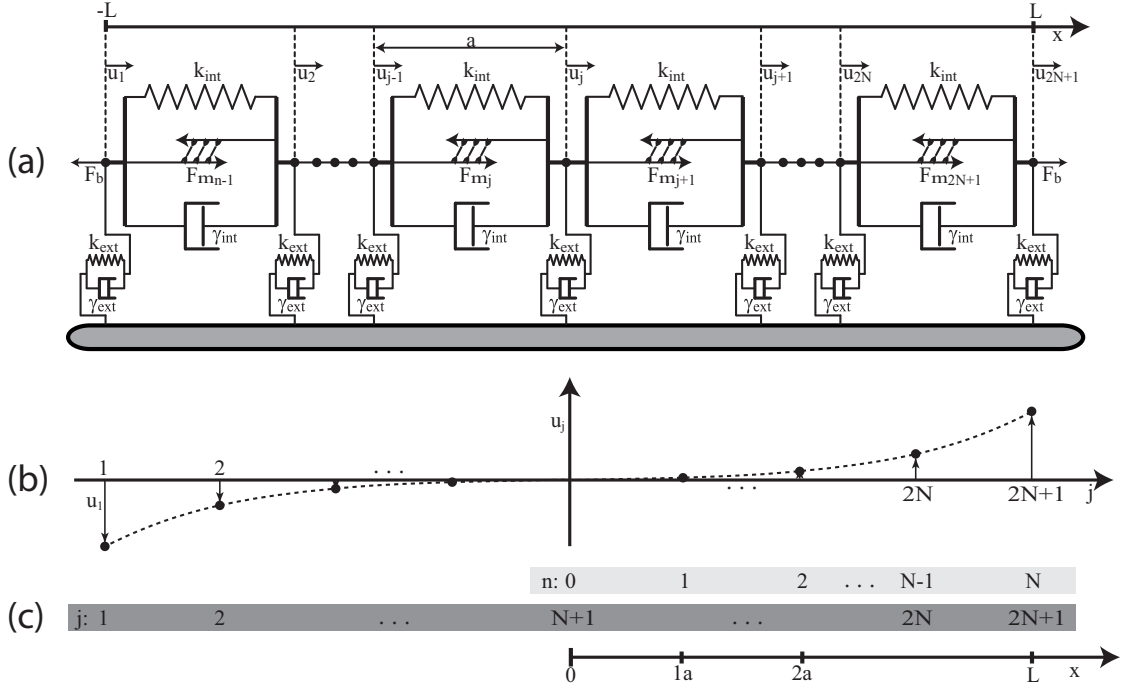}
  \caption{Discretized model. (a) The filament bundle is
modeled as a linear chain of Kelvin-Voigt bodies, each characterized by a
spring of stiffness $k_{int}$, a dashpot of viscosity
$\gamma_{int}$ and a linear extension $a$. Actomyosin contractility is described by a contractile
element with contraction force $F_m$ added to each Kelvin-Voigt body
in parallel. Viscoelastic interactions between fiber elements and their
surrounding are described by an additional set of external
Kelvin-Voigt bodies with stiffness $k_{ext}$ and viscosity
$\gamma_{ext}$. The total fiber length is $L$. $u_n$ denotes the
displacement of the $n$-th node. (b) Schematic drawing of the solution
for the node displacements assuming that both ends are pulled by an
external force $F_b>F_s$. Since both terminating nodes are pulled
outward, the solution for the displacements is antisymmetric
with respect to the center node at $n=0$, which therefore does not move.
Thus we obtain the boundary conditions
of interest, clamped at $n=0$ and pulled
by $F_b$ at $n=N$. (c) The index $n$ starts counting at the center node. The index
$j$ starts counting at the node which terminates the fiber at the
left.}
\label{fig_discrete_model}
\end{figure}

In order to find a closed analytical solution for the general stress
fiber equation, \eq{eq_model_nondim}, we discretize our model in
space.  In order to implement the correct boundary conditions at $x =
0$, it is convenient to symmetrize the system. Thus we consider a
doubled model with $2N$ units and $2N+1$ nodes as shown in
\fig{fig_discrete_model}. Like in the continuum model, internal and
external stress relaxation are modeled as Kelvin-Voigt-like, that is
($k_{int}$, $\gamma_{int}$) and ($k_{ext}$, $\gamma_{ext}$) are the spring
stiffness and the viscosity of the internal and external Kelvin-Voigt
elements, respectively. Each internal Kelvin-Voigt body is also subject
to the contractile actomyosin force $F_{m_n}$ modeled by a linearized
force-velocity relationship \cite{Howard01}:
\begin{equation} \label{eq_FVR_in_udot}
F_{m_{n}}=F_s(1-\frac{v_n}{v_0})=F_s+\frac{F_s}{v_0}(\dot{u}_n-\dot{u}_{n-1})\ .
\end{equation}
$F_{m_n}$ is the force exerted by the $n$-th motor moving with
velocity $v_n$. $v_0$ is the zero-load or maximum motor velocity and
$F_s$ is the stall force of the motor. In the final relation we have
used that the contraction velocity of the $n$-th motor, $v_n$, can be
related to the rate of elongation of the n-th sarcomeric unit
as $v_n=-(\dot{u}_n-\dot{u}_{n-1})$. 

Our model resembles the Kargin-Slonimsky-Rouse (KSR) model for
viscoelastic polymers
\cite{Kargin48,Kargin49,Rouse53,VinogradovMalkin80}, although it is
more complicated due to the presence of active stresses and the
elastic coupling to the environment. The main course of our derivation
of the solution for the discrete model follows a similar treatment
given before for the KSR-model \cite{Gotlib53}.  The force balance at
each node ${j=1,\ldots,2N+1}$ of the fiber as shown in
\fig{fig_discrete_model} reads
\begin{equation}
\label{eq_sys_non_dim}
 \begin{array}{lc}
\textrm{For~}j=1\textrm{:} & (\dot{u}_{2}-\dot{u}_1)-\Gamma\dot{u}_1+(u_{2}-u_1)-\kappa u_1 =-f(t)\\
\textrm{For~}j=2,\ldots,2N\textrm{:}\hspace{0.5cm} & (\dot{u}_{j+1}-2\dot{u}_j+\dot{u}_{j-1})-\Gamma\dot{u}_j+(u_{j+1}-2u_j+u_{j-1})-\kappa u_j=0\\
\textrm{For~}j=2N+1\textrm{:} & -(\dot{u}_{2N+1}-\dot{u}_{2N})-\Gamma\dot{u}_{2N+1}-(u_{2N+1}-u_{2N})-\kappa u_{2N+1}=f(t)\ . \\
 \end{array}
\end{equation}
We non-dimensionalized time using the time
scale $\tau$, introduced the non-dimensional parameters ($\kappa$,
$\Gamma$) and combine all inhomogeneous
boundary terms in the function $f(t)$:
\begin{equation}
\label{eq_def_Gamma_kappa}
	\tau=~\frac{v_0\gamma_{int}+F_s}{v_0 k_{int}},\hspace{1cm} \kappa=\frac{k_{ext}}{k_{int}}~,\hspace{1cm}
   \Gamma=\frac{v_0\gamma_{ext}}{v_0\gamma_{int}+F_s}~,\hspace{1cm} 
	f(t)=\frac{F_s-F_b(t)}{k_{int}}\ .
\end{equation}
It is important to note that \eq{eq_sys_non_dim} is not
made non-dimensional in regard to space; this will be done later
when the continuum limit is performed.

By taking the difference of subsequent equations in
\eq{eq_sys_non_dim} and by introducing the relative coordinates
$y_j=u_{j+1}-u_j$, we can write
\begin{equation}
\label{eq_inhom}
    \emph{\textbf{M}}_{visc}\dot{\vec{y}}+\emph{\textbf{M}}_{elas}\vec{y}=\vec{f}(t)
\end{equation}
with the $2N\times 2N$ matrix:
\begin{equation}
\label{eq_Mvisc}
 \emph{\textbf{M}}_{visc}=  \left(
 \begin{array}{cccl}
 2+\Gamma&	-1				&	0				&\cdots\\
 -1		&	2+\Gamma		&	-1				&\cdots\\
 0			&	-1				&	2 +\Gamma	&\cdots\\
 \vdots	&	\vdots		&	\vdots		&\ddots\\
 \end{array}
 \right)\ .
\end{equation}
The matrix $ \emph{\textbf{M}}_{elas}$ has the same form as
$\emph{\textbf{M}}_{visc}$, except that $\kappa$ replaces
$\Gamma$. In addition we have defined the $2N$-dimensional vectors:
\begin{equation}
\label{eq_vec_y_vec_F}
 \vec{y}(t)=   \left(
                   \begin{array}{c}
                      y_1  \\
                      y_2   \\
                      \vdots\\
                      y_{2N-1}   \\
                      y_{2N}  \\
                   \end{array}
            \right)
\hspace{0.5cm} \text{and} \hspace{0.5cm} \vec{f}(t)=   \left(
                   \begin{array}{c}
                      -f(t)  \\
                        0   \\
                      \vdots\\
                        0   \\
                      -f(t)  \\
                   \end{array}
            \right)\ .
\end{equation}

We first solve the homogeneous equation.  Let $\lambda_l$ be an
eigenvalue and let $\vec{v}_l$ be the associated eigenvector that
solves the eigenvalue problem:
\begin{equation}
\label{eq_eigenvalue_prob}
    (\emph{\textbf{M}}_{elas}-\lambda_l\emph{\textbf{M}}_{visc})\vec{v}_l=0\ .
\end{equation}
Then the general solution of the homogeneous equation is given by:
\begin{equation}
\label{eq_hom_solution}
    \vec{y}(t)=\sum_{l=1}^{2N}c_l\vec{y}_l(t)=\sum_{l=1}^{2N}c_l \vec{v}_l e^{-\lambda_l t}
\end{equation}
with the eigenvalues and eigenvectors
\begin{equation}
\label{eq_EW_n_EV}
    \lambda_l=\frac{\kappa+4 \sin^2 (\frac{\pi l}{2(2N+1)})}{\Gamma+4 \sin^2 (\frac{\pi l}{2(2N+1))}}
     \hspace{0.3cm} \text{and} \hspace{0.3cm}
     \vec{v}_l=   \left(
                   \begin{array}{c}
                      \sin (\frac{\pi  l}{2N+1})      \\[0.4cm]
                      \sin (\frac{\pi 2l}{2N+1})      \\[0.4cm]
                      \sin (\frac{\pi 3l}{2N+1})      \\
                      \vdots\\
                      \sin (\frac{\pi 2Nl}{2N+1})      \\
                   \end{array}
            \right)\ .
\end{equation}
It is straight forward to check that \eq{eq_EW_n_EV} is indeed the
solution to the eigenvalue problem defined by \eq{eq_eigenvalue_prob},
see the appendix. There we also prove that the $2N$ eigenvalues are
distinct, positive and non-zero, and that the eigenvectors are
orthogonal and their length is given by $v_l=\sqrt{(2N+1)/2}$. These
results validate the form of the homogeneous solution given in
\eq{eq_hom_solution}.

In order to determine the solution of the
inhomogeneous equation, \eq{eq_inhom}, we use variation of the coefficients:
\begin{equation}
\label{eq_inhom_ansatz}
 \vec{y}(t)=\sum_{l=1}^{2N}c_l(t)\vec{v}_l e^{-\lambda_l t}\ .
\end{equation}
Inserting this ansatz into the inhomogeneous
\eq{eq_inhom} and using the homogeneous solution yields $2N$
conditions defining the coefficients $c_l(t)$:
\begin{equation}
 \sum_l\dot{c}_l(t)\emph{\textbf{M}}_{visc}\vec{v}_l e^{-\lambda_l t}=\vec{f}(t)\ .
\end{equation}
Evaluation of the product $\emph{\textbf{M}}_{visc}\vec{v}_l$,
rewriting the $2N$ equations by components, and applying appropriate
addition theorems yields
\begin{equation}
\label{eq_determine_cldot}
 \sum_{l=1}^{2N} \dot{c}_l(t)\sin (\frac{\pi l
 j}{2N+1}) \left(\Gamma+4\sin^2 (\frac{\pi l}{2(2N+1)}) \right)e^{-\lambda_l t}=f_j(t)\ .
\end{equation}
Here the first sinus term is simply the $j$-th component of the $l$-th eigenvector.
We define a new $2N\times 2N$ matrix
\begin{equation}
\label{eq_def_U}
\textbf{U}_{j,l} = \sqrt{\frac{2}{2N+1}}\sin (\frac{\pi l j}{2N+1})
\end{equation}
and a new $2N$-dimensional vector $\vec{b}$
\begin{equation}
\label{eq_def_bl}
b_l(t)=\sqrt{\frac{2N+1}{2}}\dot{c}_l(t)\left(\Gamma+4\sin^2 (\frac{\pi l}{2(2N+1)}) \right)e^{-\lambda_l t}\ .
\end{equation}
With these definitions, \eq{eq_determine_cldot} can be rewritten as:
\begin{equation}
\label{eq_cond_on_vecb}
 \emph{\textbf{U}}\,\vec{b}(t)=\vec{f}(t)\ .
\end{equation}
Because $\emph{\textbf{U}}$ is built up by the normalized and
orthogonal eigenvectors, $\emph{\textbf{U}}^T\emph{\textbf{U}}=I$. Moreover
it is symmetric, thus $\emph{\textbf{U}}=\emph{\textbf{U}}^T=\emph{\textbf{U}}^{-1}$.
Therefore
\begin{equation}
 \vec{b}(t)=\emph{\textbf{U}}\vec{f}(t)
\end{equation}
The only non-zero components of $\vec{f}(t)$ are
$f_1=f_{2N}=-f(t)$. Therefore the solution for $\vec{b}$ is given by:
\begin{equation}
\label{eq_sol_for_bl}
 \begin{array}{rcl}
  \displaystyle b_l(t)&\displaystyle =&\displaystyle -f(t)\sqrt{\frac{2}{2N+1}}\left(\sin (\frac{\pi l}{2N+1}) + \sin (\frac{\pi 2N l}{2N+1}) \right)\\[0.5cm]
     &\displaystyle =&\displaystyle -f(t)\sqrt{\frac{2}{2N+1}} \left(1+(-1)^{l+1}\right)\sin (\frac{\pi l}{2N+1})\ .
 \end{array}
\end{equation}
We conclude that all even-numbered components of $\vec{b}$ vanish. The
coefficients $c_l(t)$ are obtained by using \eq{eq_sol_for_bl} and integrating \eq{eq_def_bl}:
\begin{equation}
 c_l(t)=\left\{
    \begin{array}{cl}
      0 &\,\,\,\,\,\,\,\,\text{if}\,\,\,l\,\,\text{even}\\[0.6cm]
      \displaystyle-\frac{4}{2N+1}\,\frac{\sin (\frac{\pi l}{2N+1}) }{\Gamma+4\sin^2 (\frac{\pi
      l}{2(2N+1)}) }\int_0^t f(t')e^{\lambda_l t'}dt'&\,\,\,\,\,\,\,\,\text{if}\,\,\,l\,\,\text{odd}
    \end{array}
    \right.
\end{equation}
The solution for the relative coordinates follows from \eq{eq_inhom_ansatz}:
\begin{equation}
\label{eq_sol_rel_coord}
 y_j(t)=-\frac{4}{2N+1}\sum_{l=1,3,5,\ldots}^{2N}\frac{\sin (\frac{\pi l}{2N+1}) \sin (\frac{\pi lj}{2N+1}) }{\gamma_{ext}+4\tilde{\gamma}_{int}\sin^2 (\frac{\pi
      l}{2(2N+1)}) }\int_0^t f(t')e^{-\lambda_l(t-t')}dt'\ .
\end{equation}
The actual displacements $u_j(t)$ are recovered from the relative coordinates by evaluating the telescoping sum:
\begin{equation}
\label{eq_u_from_y}
 \begin{array}{rcccccccc}
  u_{2N+1}-u_1 &=&\underbrace{(u_{2N+1}-u_{2N})}&+&\underbrace{(u_{2N}-u_{2N-1})}&+&\ldots &+&\underbrace{(u_2-u_1)} \\
               &=&            y_{2N}            &+&              y_{2N-1}        &+&\ldots &+&    y_{1}              \\[0.4cm]
               &=&\displaystyle\sum_{j=1}^{2N}y_j\ .
 \end{array}
\end{equation}
Since the solution has to be antisymmetric with respect to the center
node at $j=N+1$, compare \fig{fig_discrete_model}, it must hold true
that $u_{2N+1}=-u_1$ and more generally $u_{2N+1-k}=-u_{1+k}$, such
that for $k=0,\ldots,N-1$, the displacements are given by:
\begin{equation}
 \begin{array}{rcl}
  \displaystyle u_{2N+1-k}&\displaystyle =&\displaystyle \frac{1}{2}\sum_{j=1+k}^{2N-k}y_j\\[0.7cm]
                          &\displaystyle =&\displaystyle \frac{1}{2}\sum_{j=1}^{2N-k}y_j-\frac{1}{2}\sum_{j=1}^{k}y_j\\[0.7cm]
                          &\displaystyle =&\displaystyle \frac{1}{2}\sum_{l=1,3,5,\ldots}^{2N}c_l(t)e^{-\lambda_l t}\left(\sum_{j=1}^{2N-k}\sin (\frac{\pi j l}{2N+1})
                             - \sum_{j=1}^{k}\sin (\frac{\pi j l}{2N+1}) \right)\ . \\[0.7cm]
 \end{array}
\end{equation}
In the last step, we used the solution for the relative coordinates
given by \eq{eq_sol_rel_coord} and have subsequently reversed the
order of summation in both terms. The two sums in parenthesis can be
further simplified by using the identity
\begin{equation}
\label{eq_sum_over_sin} \sum_{j=0}^n\sin(j\alpha)=\frac{1}{2}\left(\cot(\alpha/2)-\frac{\cos(\alpha(n+1/2))}{\sin(\alpha/2)}\right)\ .
\end{equation}
Rewriting the result to the index $1 \le n \le N$, see \fig{fig_discrete_model}, we obtain
the desired solution of the discrete model:
\begin{equation}
\label{eq_discrete_solution_timedependent_F}
  u_{n}(t)=-\frac{2}{2N+1}\sum_{m=1}^{N}\frac{(-1)^{m-1}}{\sin\frac{\pi (2m-1)}{2(2N+1)}}\,\,\frac{\sin \frac{\pi n(2m-1)}{2N+1}\sin \frac{\pi (2m-1)}{2N+1}}{\Gamma+4\sin^2\frac{\pi
      (2m-1)}{2(2N+1)}}\int_0^t f(t')e^{-\frac{t-t'}{\tau_{m,N}}}dt'
\end{equation}
with the retardation times:
\begin{equation}
\label{eq_retardation_times_discrete}
  \tau_{m,N}=\frac{1}{\lambda_{2m-1,N}}=\frac{\Gamma+4\sin^2\frac{\pi (2m-1)}{2(2N+1)}}{\kappa+4 \sin^2\frac{\pi(2m-1)}{2(2N+1)}}\ .
\end{equation}
Note that \eq{eq_discrete_solution_timedependent_F}
gives the correct result $u_{0}=0$ for the left boundary. For this
reason, we can extend the range of validity of
\eq{eq_discrete_solution_timedependent_F} to $0 \le n \le N$. We also note
that the retardation times depend on the number of units $N$ because the
solution describes the movement of a fiber with $N$ units
which is attached at its left end $n=0$ and is pulled at its right end
$n=N$ with boundary force $f_b(t)$. It is straight forward to
confirm the validity of the derived discrete solution,
\eq{eq_discrete_solution_timedependent_F} and
\eq{eq_retardation_times_discrete}, by inserting it into the
discrete model equation, \eq{eq_sys_non_dim}.

\section{Continuum limit of the discretized model}
\label{sec_cont_limit} 

The discrete stress fiber model can be transformed to a continuum
equation by considering the limit $N\rightarrow\infty$ while the length $L$ of the fiber is
kept constant. In this process, the stress fiber length $L$ is
subdivided into incremental smaller pieces of length
$a_N=L/N$. Thereby it has to be ensured that the effective
viscoelastic properties of the whole fiber are conserved. This is
accomplished by re-scaling all viscoelastic constants in each
iteration step with the appropriate scaling factor
$\phi_N=\frac{a}{a_N}=\frac{Na}{L}$ according to:
\begin{equation}
\label{eq_scaling_const}
 \begin{array}{ccccccc}
  k_{N,int} & = & \phi_N k_{int}                         &\,\,\,\,\,\,\text{and}\,\,\,\,\,\,& \gamma_{N,int} & = &  \phi_N\gamma_{int}\\[0.2cm]
  k_{N,ext} & = & {\displaystyle \phi_N^{-1}k_{ext}}  &\,\,\,\,\,\,\text{and}\,\,\,\,\,\,& \gamma_{N,ext} & = &  {\displaystyle\phi_N^{-1}\gamma_{ext}}
 \end{array}
\end{equation}
To further clarify this procedure consider a single harmonic spring of
resting length $a$ and stiffness $k$. This spring is equivalent to two
springs of length $a/2$ and stiffness $2k$ that are connected in
series. Here, the scaling factor is $\phi_2=\frac{a}{a_2}=2$. Thus,
the stiffness $k_{N,int}$ in \eq{eq_scaling_const} represents the
stiffness of a fiber fragment of length $a_N$ and increases linearly
with the number of partitions $N$, whereas $k_{int}$ is the reference
stiffness of a fiber fragment of length $a$. A typical value for the
length scale $a$ would be $a=1\,\mu\textrm{m}$, the typical length of
sarcomeric units in stress fibers \cite{uss:colo09}. While
$k_{N,int}$ increases linearly with the number of partitions,
$k_{N,ext}$ decreases according to $1/N$. Similarly it follows that
the viscous parameter $\gamma_{N,int}$ and $\gamma_{N,ext}$ scale as
$k_{N,int}$ and $k_{N,ext}$, respectively. The
non-dimensional parameters and the boundary force scale like:
\begin{equation}
\label{eq_scaling_nondim_const}
  \Gamma_N=\phi_N^{-2} \Gamma, \hspace{1cm} \kappa_N=\phi_N^{-2}\kappa, \hspace{1cm} f_N(t)=\phi_N^{-1}f(t)\ .
\end{equation}
We begin the limiting procedure by introducing the continuous spatial
variable $x=n a_N$, denoting the position of the $n$-th node within
the discrete chain with $N$ units. Then
\eq{eq_sys_non_dim} yields (also compare \fig{fig_discrete_model}):
\begin{equation}
\label{eq_sys_non_dim_n_counting}
 \begin{array}{l}
\textrm{For~}n=0:\\
 u(0) =0\\[0.3cm]
\textrm{For~}n=1,\ldots,N\textrm{:}\hspace{0.5cm} \\ \dot{u}(x+a_N)-2\dot{u}(x)+\dot{u}(x-a_N)-\Gamma_N\dot{u}(x)+u(x+a_N)-2u(x)+u(x-a_N)-\kappa_N u=0\\[0.3cm]
\textrm{For~}n=N\textrm{:} \\ \dot{u}(L)-\dot{u}(L-a_N)+\Gamma_N\dot{u}(L)+u(L)-u(L-a_N)+\kappa_Nu(L)+f_N(t)=0\\
 \end{array}
\end{equation}
Using the scaling relations for the viscoelastic parameters given in
\eq{eq_scaling_nondim_const} and conducting the limit
$N\rightarrow\infty$ yields for $n=1,\ldots, N$:
\begin{equation}
 \begin{array}{lcl}
    \displaystyle a^2\lim_{N\rightarrow\infty}\left(\frac{\dot{u}(x+a_N)-2\dot{u}(x)+\dot{u}(x-a_N)}{{a_N}^2}\right)-\Gamma\dot{u}(x)&\displaystyle +\ldots&\\[0.7cm]
    \displaystyle a^2\lim_{N\rightarrow\infty}\left(\frac{u(x+a_N)-2u(x)+u(x-a_N)}{{a_N}^2}\right)-\kappa u(x)&\displaystyle = & \displaystyle 0
 \end{array}
\end{equation}
Since $a_N$ is a sequence which converges to zero, the limits define
the second derivative of $u$ with respect to $x$. The continuum limit
of the upper equation results in a partial differential equation for
the displacement $u(x,t)$. The highest order term will contain mixed
derivatives in $x$ and $t$, namely, $\partial_x^2\dot{u}$. Similarly,
the limiting process can be performed for the boundary condition at
the right end. Note that at this point the spatial variable evaluates
to $x=Na_N=L$:
\begin{equation}
 \begin{array}{ll}
    \displaystyle a\lim_{N\rightarrow\infty}\left(\frac{\dot{u}(L)-\dot{u}(L-a_N)}{a_N}\right)+\Gamma\lim_{N\rightarrow\infty}\frac{a_N}{a}\dot{u}(L)+\ldots&\\[0.7cm]
    \displaystyle a\lim_{N\rightarrow\infty}\left(\frac{u(L)-u(L-a_N)}{a_N}\right)+\kappa\lim_{N\rightarrow\infty}\frac{a_N}{a}u(L)  +f(t)=&\displaystyle 0
 \end{array}
\end{equation}
In each line of the equation, the first limit gives the first
derivative of $u$ with respect to $x$ evaluated at $x=L$ and the
second limit in each line vanishes as $a_N$ converges to
zero. Consequently, in the continuum representation, the stresses
which originate from shearing the environment cannot contribute to the
boundary condition. Our continuum model for stress fibers
defined by \eq{eq_model_nondim} and \eq{eq_model_nondim_BC} is
recovered after non-dimensionalizing $x,L,u,f$ using the typical
length scale $a$.

To obtain a closed solution for the continuous model, we apply the
continuum limit to the discrete model solution. The limiting procedure
is first performed on the retardation times of the discrete model
given by \eq{eq_retardation_times_discrete}:
\begin{equation}
\label{eq_retardation_times_limit}
  \tau_{m,N}=\frac{\Gamma_{N}+4\sin^2\frac{\pi (2m-1)}{2(2N+1)}}{\kappa_{N}+4 \sin^2\frac{\pi(2m-1)}{2(2N+1)}}
=\frac{\Gamma l^2+4N^2\sin^2\frac{\pi (2m-1)}{2(2N+1)}}{\kappa l^2+4 N^2\sin^2\frac{\pi (2m-1)}{2(2N+1)}}\ .
\end{equation}
Performing the limit $N\rightarrow\infty$ yields the retardation times of the continuum model:
\begin{equation}
\label{eq_retardation_times_cont}
  \tau_m=\frac{4\Gamma l^2+\pi^2(2m-1)^2}{4\kappa l^2+\pi^2(2m-1)^2}\ .
\end{equation}
Since $1\leq m\leq\infty$, the upper relation defines infinitely
many discrete retardation times, non-dimensionalized by $\tau$. From \eq{eq_retardation_times_cont} we deduce that the retardation times
are bounded by the extreme values $\tau_{1}$ and $\tau_{\infty}$
according to:
\begin{equation}
\label{eq_range_spec}
  1\leq\tau_m\leq\frac{4\Gamma l^2+\pi^2}{4\kappa l^2+\pi^2} \,\,\,\,\,\, \text{or}  \,\,\,\,\,\, \frac{4\Gamma l^2+\pi^2}{4\kappa l^2+\pi^2}\leq\tau_m\leq 1\ .
\end{equation}
The first relation holds if $\kappa\leq\Gamma$, whereas the second
holds if $\kappa\geq\Gamma$. In the special case $\kappa=\Gamma$ the
finite range of possible values collapses to the single retardation
time $\tau$ which leads to a very simple form of the analytical
solution as shown above with \eq{eq_solution_kappa_eq_Gamma}.

The limiting procedure applied in \eq{eq_retardation_times_limit} can
be carried out similarly on the remaining $N$-dependent terms of the
discrete solution given by \eq{eq_discrete_solution_timedependent_F}.
This yields our central result, i.e. the solution for the continuous
boundary value problem defined by \eq{eq_model_nondim} and
\eq{eq_model_nondim_BC}:
\begin{equation}
\label{eq_general_solution}
 u(x,t) \displaystyle = \displaystyle 8 l
\sum_{m=1}^\infty\frac{(-1)^{m+1}\sin\frac{\pi x (2m-1)}{2l}}{4\Gamma l^2 + \pi^2(2m-1)^2} \int_0^t (f_b(t') -f_s) e^{-\frac{t-t'} {\tau_m}}dt'
\end{equation}
\eq{eq_general_solution} in combination with
\eq{eq_retardation_times_cont} is the general solution of our
continuum model. We successfully checked the validity of our
analytical solution by comparision with a numerical solution of
\eq{eq_model_nondim}. One big advantage of the analytical solution is
that it can be easily used to evaluate experimental data. In the
following, we will discuss its consequences for the two special cases
of laser cutting and cyclic loading.

\section{Laser cutting}
\label{sec_laser_cutting}

\begin{figure}[t]
        \centering
      \includegraphics[width=0.80\textwidth]{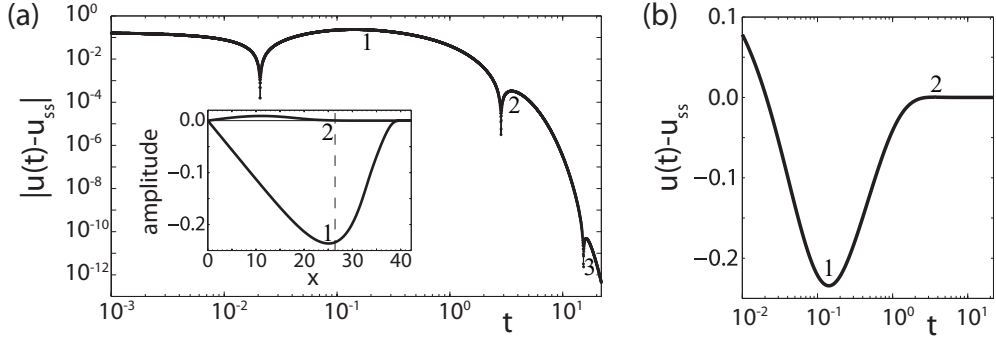}
        \caption{First oscillation. (a) If $\Gamma<\kappa$, the
displacements of inner fiber segments exhibit damped oscillations
around their final steady state $u_{ss}(x)$. Here we show a
log-log-plot of the time course of the absolute difference
$|u(x,t)-u_{ss}(x)|$ calculated from
\eq{eq_cont_solution_from_cont_limit} for the position
$x=26.4$, which is close to the position with maximal
amplitude and corresponds to the band $n=7$ in
\fig{fig_actin_dynamics}. The inset gives the amplitudes of the first and
second oscillation along the fiber, with maxima $236\,\textrm{nm}$ and
$\textrm{9~nm}$, respectively. The position $x=26.4$ used here
is highlighted as dashed line. (b) The
difference $u(x,t)-u_{ss}$ at the position $x=26.4$ is shown on a linear scale.
Numbering of the extremal values are included for comparison with (a).
Parameters for (a) and (b) are as in \fig{fig_actin_dynamics}:
$(\kappa,f_s,\Gamma)=(0.028,0.39,0)$ and $l=42.2$,
${a=1\,\mu\textrm{m}}$.
        }
        \label{fig_damped_osc_non_dim}
\end{figure}

\begin{figure}[t]
\centering
\includegraphics[width=0.80\textwidth]{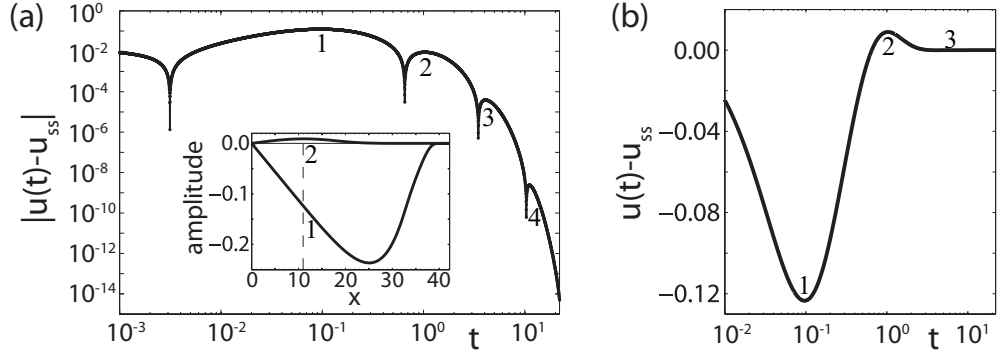}
 \caption{Second oscillation. (a) Time course of the absolute difference $|u(x,t)-u_{ss}|$
   on a log-log-scale at the position $x=11.0$, where the amplitude of the second oscillation
   attains its maximum. The inset again shows the maximum amplitude of the
   first and second oscillation along the fiber, but now the position $x=11.0$
   is highlighted as dashed line. (b) The difference $u(x,t)-u_{ss}$
   at the position $x=11.0$ is shown on a linear scale. Numbering of
   the extremal values are included for comparison with (a). Parameters used
   for (a) and (b) are the same as in \fig{fig_damped_osc_non_dim} and
   \fig{fig_actin_dynamics}.}
\label{fig_osc_at_second_max}
\end{figure}

If a fiber is cut by a train of laser pulses,
then there are no external forces acting anymore on the free fiber
end and $f_b(t)$ vanishes.  Then
\eq{eq_general_solution} can be written as:
\begin{equation}
\label{eq_cont_solution_from_cont_limit}
 u(x,t)=\sum_{m=1}^\infty S_m(x)\left(1-e^{-\frac{t}{\tau_m}}\right)\ .
\end{equation}
The solution for the displacement can be understood as a retardation
process with infinitely many discrete retardation times $\tau_m$ given
by \eq{eq_retardation_times_cont} and associated, spatially dependent
amplitudes $S_m(x)$. The amplitudes are given by:
\begin{equation}
\label{eq_spec_general_x}
 S_m(x)=8 f_s l\frac{(-1)^{m}\sin\frac{\pi x (2m-1)}{2l}}{4\kappa
 l^2+\pi^2(2m-1)^2}\ .
\end{equation}
The solution for the displacement at $x=l$, the position of the cut,
is particularly simple. At this special position, the amplitudes
have a linear relation to the corresponding retardation times:
\begin{equation}
\label{eq_spec_x_eq_L}
    S_m(l)=-\frac{2 f_s}{l} \frac{1-\tau_m}{\kappa-\Gamma}\ .
\end{equation}
Since the range of possible retardation times is bounded according to
\eq{eq_range_spec}, it follows that the spectrum at $x=l$ has only
negative amplitudes and the resulting solution for the displacement at
$x=l$ is always a monotonically decreasing function. However, this is
not true for arbitrary $x$. Inspection of \eq{eq_spec_general_x}
yields that negative as well as positive amplitudes appear
simultaneously. Since $x=l$ evaluates the numerator in
\eq{eq_spec_general_x} at its maximum, the resulting spectrum
constitutes a lower bound for the negative amplitudes of the spectra
with $x\neq l$. Similarly, the absolute value of \eq{eq_spec_x_eq_L}
gives an upper bound for all positive amplitudes.  Thus, the
retardation spectra with $x\neq l$ oscillate around zero within an
envelope for the amplitudes that decays linearly toward zero.  This
can lead to damped oscillations in the displacement of inner fiber
bands about their stationary value. An representative time course is shown
in \fig{fig_damped_osc_non_dim}. The emergence of these oscillations
is particularly interesting since the stress fiber is modeled in the
overdamped limit, that is, inertia terms are neglected. We find that
these damped oscillations in this inertia-free system occur only for
$\Gamma/\kappa < 1$, but then for all positions $x \neq l$. The
amplitude of these oscillations reach their maximum at distinct
positions along the fiber, as shown by the inset to
\fig{fig_damped_osc_non_dim}. The location of the maxima moves further
away from the cut (toward smaller $x$-values) with increasing order of
the oscillation. \fig{fig_damped_osc_non_dim} shows the time course of
the displacement at $x=26.4$ which is close to the position where the
first oscillation reaches its maximum. Using the same parameters as in
\fig{fig_damped_osc_non_dim}, we show in \fig{fig_osc_at_second_max}
the time course of the displacement at $x=11.0$ where the second
oscillation reaches its maximum. Since the oscillations are strongly
damped, the maximum amplitude of the oscillations also decreases with
the order. While the maximal amplitude of the first oscillation can
reach hundreds of nanometers ($236\,\textrm{nm}$ at $x=24.9$, see
inset to \fig{fig_damped_osc_non_dim}), the maximal amplitude of the
second oscillation is already much smaller and only of the order of
tens of nanometers ($9\,\textrm{nm}$ at $x=11.0$, see
\fig{fig_osc_at_second_max}). Thus, in order to detect the
oscillations in experiments, it is essential to measure close to where
the oscillations reach their respective maximum amplitude. We
demonstrate this by showing predicted time courses of the difference
$u(x,t)-u_{ss}$ at the two positions in
\fig{fig_damped_osc_non_dim}~(b) and \fig{fig_osc_at_second_max}~(b),
respectively. While the first oscillation is most prominent in
\fig{fig_damped_osc_non_dim}~(b), the second oscillations is not
detectable at this position. In contrast, the amplitude of the first
oscillation in \fig{fig_osc_at_second_max}~(b) is reduced compared to
\fig{fig_damped_osc_non_dim}~(b) but the amplitude of the second
oscillation is much larger and becomes detectable.

\begin{figure}[ht]
\centering
\includegraphics[width=0.6\textwidth]{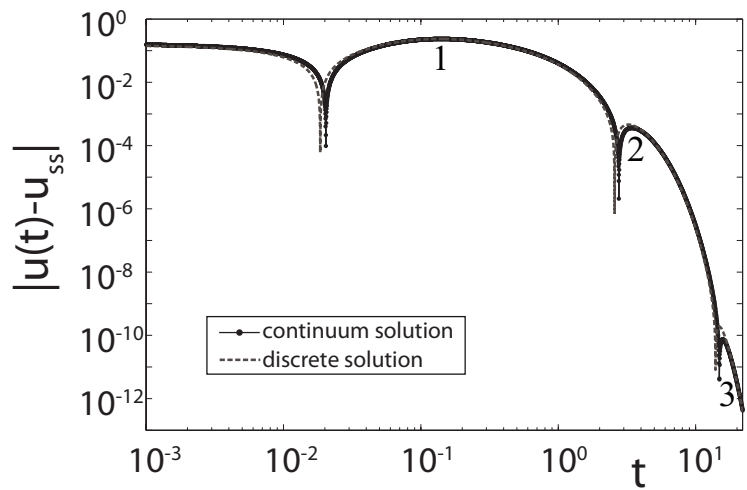}
 \caption{Comparision of $|u(x,t)-u_{ss}(x)|$ for the continuum model (solid)
and for the discrete model (dashed). The continuum solution
is calculated for a fiber of length $l=42$ at position $x=26$. The
discrete solution is calculated for a fiber with $N=42$ subunits at
node $n=26$. Both solutions were calculated for the same parameters
$(\kappa, \Gamma, f_s)=(0.028,0,0.39)$ as extracted from experimental data.
The two solutions agree well with each other and both
models predict oscillations.}
\label{SFig1_compare_disc_cont}
\end{figure}

To show that the oscillations are not an artifact introduced by the
continuum limit, we have compared solutions of corresponding
continuous and discrete models. Results are shown in
\fig{SFig1_compare_disc_cont} where we have used the same parameters
as for \fig{fig_damped_osc_non_dim} and
\fig{fig_osc_at_second_max}. To facilitate comparison between
continuum and discrete model, solutions are calculated for integer
fiber lengths and integer positions. We find that continuum and
discrete solution agree very well and, most importantly, both predict
oscillations when $\Gamma<\kappa$.

\begin{figure}[t] 
        \centering
        \includegraphics[width=0.90\textwidth]{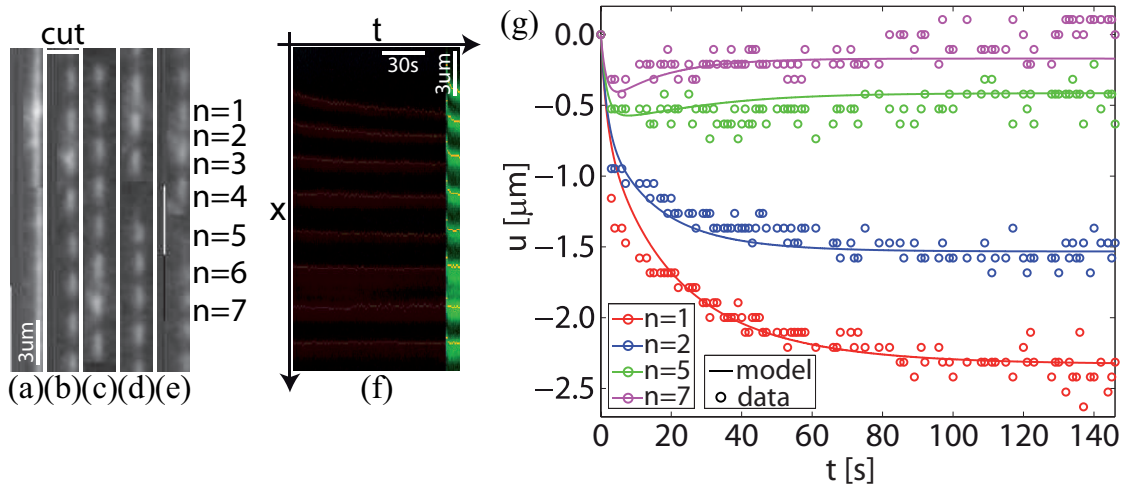}
        \caption{(Color online) Experimental results. (a) GFP-actin stress fiber prior to
patterning by photo bleaching, scale bar: $3~\mu\textrm{m}$. (b)
Fiber bleached with stripe pattern. (c-e) Stress fiber $1\,\textrm{s}$, $30\,\textrm{s}$ and
$140\,\textrm{s}$ after laser cutting. (f) Time-space kymograph
reconstructed from fluorescence intensity profiles along the stress
fiber. Band positions are extracted by edge-detection (solid lines),
scale bars: $30~\textrm{s}$ and $3~\mu\textrm{m}$.
(g) Model fit to the displacement data of shown bands (n increasing from bottom to top) 
with initial positions $x_{n=1,2,5,7}= (42.2, 39.7, 31.8, 26.4)\mu\textrm{m}$
yields $(\kappa,f_s,\tau,\Gamma)=(0.028,0.39,34\,\textrm{s},0.0)$ with
$a=1.0\,\mu\textrm{m}$. Note that $\Gamma/\kappa\ll 1$ and
that the experimental data provide evidence for the predicted oscillations,
because the curves for n=5 and 7 show dips at 8s and 5s after cutting, respectively. 
        }
        \label{fig_actin_dynamics}
\end{figure}

Because of the analytical solution
\eq{eq_cont_solution_from_cont_limit}, we can easily apply our model
to evaluate experimental data for stress fiber contraction dynamics
induced by laser nano-surgery \cite{uss:colo09}. Briefly, Ptk-2 cells
were tranfected with GFP-actin and a stripe pattern was bleached into
their stress fibers. 10 s later the stress fibers were cut with a
laser and their retraction was recorded over several
minutes. Kymographs were constructed and for each band, the retraction
trace was extracted by edge detection. Least-square fitting of the
theoretical predictions to four selected bands simultaneously was used
to estimate the four model parameters ($\kappa, f_s, \tau,
\Gamma$). An representative example for the outcome of this procedure
is shown in \fig{fig_actin_dynamics} (more examples and the details of
our experiments are provided in the supplementary material). We find
that $\Gamma = (0.52\pm0.23)\cdot 10^{-3}\ll 1$ and $\Gamma/\kappa =
0.013\pm0.021 \ll 1$ (mean $\pm$ std, $N=6$). This means that in our
experiments the second relation of \eq{eq_range_spec} applies and that
the oscillations demonstrated by \fig{fig_damped_osc_non_dim} are
predicted for this experimental system. For the positions
corresponding to bands $n=5$ and $n=7$ our model predicts minima at
$8\,\textrm{s}$ and $5\,\textrm{s}$, respectively.  Indeed these
minima appear as dips in the experimental data shown in
\fig{fig_actin_dynamics}(g).

\section{Cyclic loading}
\label{sec_cyclic_loading}
 
The response of stress fibers to cyclic loading is characterized by
the complex modulus which we derive from the general solution
\eq{eq_general_solution} by assuming a cyclic boundary force
$f_b(t)=f_s+f_0 e^{i\omega t}$, with a constant offset compensating
the stall force of the molecular motors. Evaluation of the resulting
integral in \eq{eq_general_solution} yields:
\begin{equation}
\label{eq_cont_solution_cyclic}
 u(x,t)=8 l f_0 \sum_{m=1}^\infty\frac{(-1)^{m+1}\sin\frac{\pi x (2m-1)}{2l}}{4\kappa l^2+\pi^2(2m-1)^2} \,\frac{1}{i\omega\tau_m+1}\left(e^{i\omega t}-e^{-\frac{t}{\tau_m}}\right)\ .
\end{equation}
Inspection of the time-dependent terms yields that the solution for
the displacements approaches a harmonic oscillation. The deviations
decay exponentially in time, according to $e^{-t/\tau_m}$. As a
consequence, in the limit for large times, the fiber displacements
also oscillate with the same frequency $\omega$ as the force input,
but the stationary phase shift between displacements $u(x,t)$ and
$f_b(t)$ might vary spatially along the fiber. In the following, we
are only interested in the response of the fiber as a whole, i.e. we
focus on the displacement at $x=l$. With the above arguments, we
find in the limit for large times:

\small
\begin{equation}
\label{eq_sol_cyclic}
\begin{array}{rcc}
 \displaystyle u(l,t)=  & \displaystyle \underbrace{\sum_{m=1}^\infty\frac{8 l}{4\kappa l^2+ \pi^2(2m-1)^2}\,\frac{1}{i\omega\tau_m+1}} \cdot& \displaystyle f_0 e^{i\omega t}\ . \\[0.5cm]
                        & \displaystyle =1/\mathcal{G}^*(\omega) &  \\
\end{array}
\end{equation}
\normalsize The complex modulus, non-dimensionalized by $K_{int}$, can
be deduced from \eq{eq_sol_cyclic} by noting that the cyclic force
input $f_0 e^{i\omega t}$ and the creep response of the fiber $u(l,t)$
are connected by the inverse of the complex modulus
\cite{b:pipkin86}. The expression for the complex modulus can be
separated into its real and imaginary part, the storage and the loss
modulus, respectively:
\begin{equation}
\label{eq_def_storage_n_loss_modulus}
\begin{array}{rcc}
    \displaystyle\mathcal{G}^*(\omega)= &\displaystyle\underbrace{\frac{p(\omega)}{p^2(\omega)+q^2(\omega)}}+   &\displaystyle i\underbrace{\frac{q(\omega)}{p^2(\omega)+q^2(\omega)}}\\[0.5cm]
                                                &\displaystyle =\mathcal{G}'(\omega)                                                                 &\displaystyle =\mathcal{G}''(\omega)\\
\end{array}
\end{equation}
with
\begin{equation}
 \begin{array}{rl}
    p(\omega)=&\displaystyle 8l\sum_{m=1}^\infty\frac{1}{4\kappa l^2+\pi^2(2m-1)^2}\,\frac{1}{\omega^2\tau^2_m+1}\ , \\[0.7cm]
    q(\omega)=&\displaystyle 8l\sum_{m=1}^\infty\frac{1}{4\kappa l^2+\pi^2(2m-1)^2}\,\frac{\omega \tau_m}{\omega^2\tau^2_m+1}\ . \\
 \end{array}
\end{equation}

\begin{figure}[t] 
        \centering
        \includegraphics[width=0.66\textwidth]{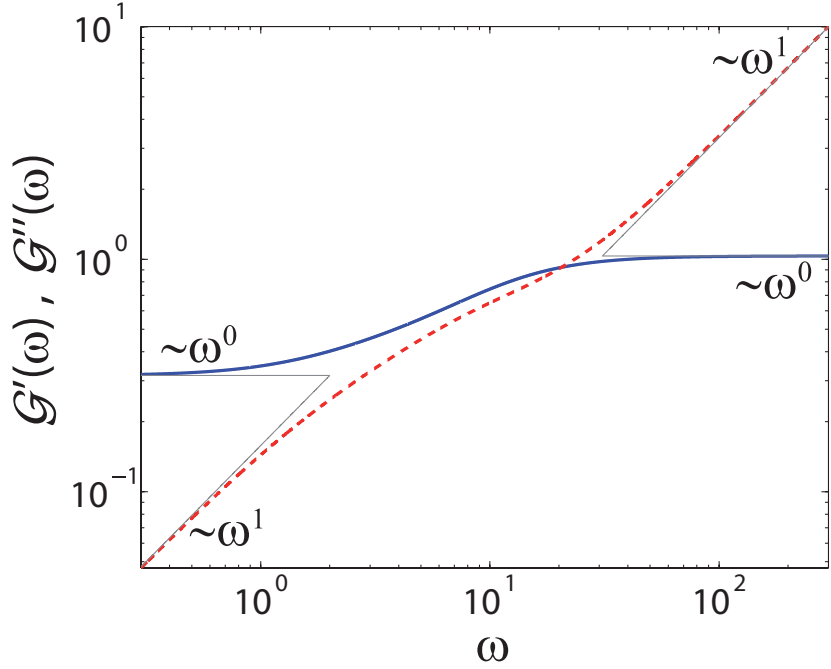}
        \caption{(Color online) Log-log plot of storage modulus
(solid) and  loss modulus (dashed). Scaling at low and high frequencies is
shown for both storage and loss modulus. Used parameters are
$(\kappa,\Gamma)=(0.1,0)$ and  $l=30$.
        }
        \label{fig_loglogGstar_L30_k01_Gamma0}
\end{figure}

An alternative, more concise expression for the complex modulus can be
derived by solving the Laplace-transformed model equation for the
situation of a sudden force application, $f_b(t)=f_s+f_0\theta(t)$,
where $\theta(t)$ is the unit step function. Solution of this
Laplace-transformed boundary value problem for
$\bar{u}(l,s)=\int_0^\infty u(l,t)e^{-st}$, with $s=\gamma+i\omega$,
directly yields the Laplace-transformed creep
compliance, $\bar{J}(s)=\bar{u}(l,s)/f_0$. It is connected to the
complex modulus by:
\begin{equation}
\label{eq_Gstar_analytic}
 \mathcal{G}^*(\omega)=\lim_{\gamma\rightarrow 0}\frac{1}{s\bar{J}(s)} = \frac{\sqrt{1+i\omega}\sqrt{i
 \Gamma \omega+\kappa}}{\tanh\left(l\frac{
 \sqrt{i \Gamma \omega + \kappa}}
 {\sqrt{1+i\omega}}\right)}\ .
\end{equation}
\eq{eq_def_storage_n_loss_modulus} or \eq{eq_Gstar_analytic}
are equivalent expressions for the complex modulus of the stress fiber
model. To further study its frequency dependence we use
\eq{eq_Gstar_analytic}. In the special case $\Gamma/\kappa=1$, it
simplifies to $\mathcal{G}^*(\omega)=(1+i\omega)\sqrt{\kappa}/
\tanh(l\sqrt{\kappa})$. The storage modulus becomes a constant, and
the loss modulus is linearly dependent on the frequency. These are the
characteristics of a Kelvin-Voigt body. The more the
ratio $\Gamma/\kappa$ differs from unity, the larger are the deviations
from these simple characteristics. To study the general case
$\Gamma/\kappa\neq1$, consider the limits $\omega\rightarrow 0$ and
$\omega\rightarrow \infty$. In both limits, the stress fiber model
again exhibits the characteristics of a Kelvin-Voigt body. The
explicit values for the limit $\omega\rightarrow 0$ are:
\begin{equation}
 \begin{array}{rl}
\displaystyle\mathcal{G}_0'=&\displaystyle\sqrt{\kappa}\coth(l\sqrt{\kappa})\\[0.5cm]
\displaystyle\mathcal{G}_0''(\omega)=&\displaystyle\frac{1}{4\kappa}\csch^2(l\sqrt{\kappa})\left(2l\kappa(\kappa-\Gamma)+\sqrt{\kappa}(\kappa+\Gamma)\sinh(2l\sqrt{\kappa})\right)\omega.\\
 \end{array}
\end{equation}
Similarly, in the limit $\omega\rightarrow \infty$, we find:
\begin{equation}
 \begin{array}{rl}
\displaystyle\mathcal{G}_{\infty}'=&\displaystyle\frac{1}{4\Gamma}\csch^2(l\sqrt{\Gamma})\left(2l\Gamma(\Gamma-\kappa)+\sqrt{\Gamma}(\Gamma+\kappa)\sinh(2l\sqrt{\Gamma})\right)\\[0.5cm]
\displaystyle\mathcal{G}_{\infty}''(\omega)=&\displaystyle\sqrt{\Gamma}\coth(l\sqrt{\Gamma})\omega.\\  
 \end{array}
\end{equation}
It holds that $\mathcal{G}_{\infty}'/\mathcal{G}_0'\geq 1$, with equality for
$\Gamma=\kappa$. A similar relation holds for the slope of the loss
modulus at high and low frequencies. \fig{fig_loglogGstar_L30_k01_Gamma0}
shows the predicted frequency dependences of $\mathcal{G}'$ and $\mathcal{G}''$.

\section{Discussion}
\label{sec_discussion}

Here we have presented a complete analytical solution of a generic
continuum model for the viscoelastic properties of actively
contracting filament bundles. Our model contains the most important
basic features which are known to be involved in the function of
stress fibers, namely internal viscoelasticity, active contractility
by molecular motors, and viscous and elastic coupling to the
environment. The resulting stress fiber equation,
\eq{eq_model_nondim}, can be solved with numerical methods for partial
differential equations. In this paper, we have shown that a general
solution can be derived by first discretizing the equation in
space. In order to implement the correct boundary conditions, the
system is symmetrized by doubling its size. The resulting system of
ordinary differential equations leads to an eigenvalue problem which
can be solved exactly, leading to
\eq{eq_discrete_solution_timedependent_F}. A continuum limit needs to
take care of the appropriate rescaling of the viscoelastic parameters
and finally leads to the general solution \eq{eq_general_solution} for
the stress fiber equation. The validity of our analytical solution has
been successfully checked by comparing it with both the discrete and
numerical solutions.

Due to their analytical nature, our results can be easily used to
evaluate experimental data.  Here we have demonstrated this for the
case of laser cutting of stress fibers. In an earlier experimental
study \cite{uss:colo09}, we focused on the movement of the first three
bands ($n=1,2,3$) of the stress fiber. These bands are within less
than $10~\mu\textrm{m}$ from the fiber tip and thus we did not report
the oscillatory feature of bands farther away from the cut. After
prediction of these oscillations by our analytical results, we
evaluated the experimental data in this respect and indeed found
evidence for their occurance (\fig{fig_actin_dynamics} and
supplementary material). This was possible with conventional light
microscopy because the amplitude of the first oscillation can reach
hundreds of nanometers. The amplitude of the second oscillation,
however, is predicted to be typically on the order of tens of
nanometers, which is below our resolution limit. In the future,
super-resolution microscopy or single particle tracking might allow a
nanometer-precise validation of our theoretical predictions.

The extracted parameter values suggest that the frictional coupling
between stress fiber and cytoplasm, quantified by $\Gamma$, is not
relevant in our experiments, and that the retraction dynamics is
dominated by the elastic foundation quantified by $\kappa$
\cite{uss:colo09}. However, the elastic coupling to the environment
might depend on cell type and substrate coating.  In fact our findings
differ from the results of an earlier study, which neglected elastic,
but predicted high frictional coupling \cite{Stachowiak09}. In our
model, high frictional coupling corresponds to $\Gamma/\kappa > 1$ and
thus no oscillations are expected in this case. It would be
interesting to cut stress fibers in cells grown on micro-patterned
surfaces that prevent substrate attachment along the fiber. We then
would expect not only the transition from elastic to viscous coupling,
but also the disappearance of the oscillations.

As a second application of our theoretical results, we suggest to
measure the viscoelastic response function $G(\omega)$ of single
stress fibers. This could be done with AFM or similar setups either on
live cells \cite{uss:colo09} or on single stress fibers extracted from
cells \cite{katoh_isolation_1998,Deguchi06,Matsui09}. In this case,
our model could provide a valuable basis for evaluating changes in the
viscoelastic properties of stress fibers induced by changes in motor
regulation, e.g.\ by calcium concentration or pharmacological
compounds.

In summary, our analytical results of a generic model open up the
perspective of quantitatively evaluating the physical properties for
any kind of contractile filament bundle. In order to apply this
approach to more complicated cellular or biomimetic systems, it would
be interesting to go beyond the one-dimensional geometry of bundles
and to also consider higher dimensional arrangements of contractile
elements
\cite{Gittes97,Gardel04,Mizuno07,koenderink_active_2009,Fabry01,Fernandez06},
which could be modeled for example by appropriately modified two- and
three-dimensional networks \cite{c:coug03,uss:paul08a,uss:bisc08a}.

\section{Acknowledgments}

EHKS and USS are members of the Heidelberg cluster of excellence
CellNetworks. USS was supported by the Karlsruhe cluster of excellence
Center for Functional Nanostructures (CFN) and by the MechanoSys-grant
from the Federal Ministry of Education and Research (BMBF) of
Germany. AB was supported by the NIH Grant R01 GM071868 and by the
German Research Foundation (DFG) through fellowship BE4547/1-1.

\section{Appendix: Proofs for eigenvalues and eigenvectors}
\label{app_proof_eigenval}

In the main text we have used the eigenvalues and eigenvectors given
by \eq{eq_EW_n_EV} without proving that this system indeed solves the
eigenvalue problem defined by \eq{eq_eigenvalue_prob}. Here, we verify
the solution to the eigenvalue problem 
and prove the following properties of the eigenvalues and eigenvectors:
\renewcommand{\labelenumi}{(\theenumi)}
\begin{enumerate}
 \item The eigenvalues are distinct, positive and non-zero.
 \item The eigenvectors are orthogonal and their length is given
 by $v_l=\sqrt{(2N+1)/2}$
\end{enumerate}
In order to verify the given eigenvalues and eigenvectors, we first
rewrite the matrix $\emph{\textbf{M}}_{elas}-\lambda_l
\emph{\textbf{M}}_{visc}$ as:
\begin{equation}
\label{eq_app_M}
    \emph{\textbf{M}}_{elas}-\lambda_l \emph{\textbf{M}}_{visc}=
                           \left(
                                    \begin{array}{cccl}
                                          2B-A  &  -B   &   0   & \cdots\\
                                           -B   &  2B-A &  -B   & \cdots\\
                                            0   &  -B   &  2B-A & \cdots\\
                                         \vdots & \vdots& \vdots& \ddots\\
                                    \end{array}
                            \right)
\end{equation}
Where $B=1-\lambda_l$ and $A=-\kappa+\lambda_l \Gamma$. By using \eq{eq_EW_n_EV} one can express $A$ in terms of $B$:
\begin{equation}
    A=4B\sin^2\frac{\pi l}{2(2N+1)}
\end{equation}
Substitution of this relation into \eq{eq_app_M} and the application of the addition theorem $\cos 2\alpha=1-2\sin^2\alpha$ yields:
\begin{equation}
\begin{array}{lcl}
    \emph{\textbf{M}}_{elas}-\lambda_l \emph{\textbf{M}}_{visc}=&
                           B\left(
                                    \begin{array}{cccl}
                                          2\cos\left(\frac{\pi l}{2N+1}\right)  &  -1   &   0      &\cdots\\
                                           -1   &  2\cos\left(\frac{\pi l}{2N+1}\right) &  -1      &\cdots\\
                                            0   &  -1   &  2\cos\left(\frac{\pi l}{2N+1}\right)    &\cdots\\
                                         \vdots & \vdots& \vdots& \ddots\\
                                    \end{array}
                            \right)&
=:B \emph{\textbf{M}}_l
\end{array}
\end{equation}
To prove \eq{eq_eigenvalue_prob} it has to be shown that the product
$\emph{\textbf{M}}_l\vec{v}_l$ vanishes for all $l=1,\ldots,2N$. The
$m$-th component of the vector which results from this product is
given below. It simplifies to zero after application of the addition
theorem $2\cos\alpha \,
\sin\beta=\sin(\alpha+\beta)+\sin(\beta-\alpha)$:
\begin{equation}
    -\sin\left(\frac{\pi (m-1)l}{2N+1}\right)+2\cos\left(\frac{\pi l}{2N+1}\right)\sin\left(\frac{\pi ml}{2N+1}\right)-\sin\left(\frac{\pi
    (m+1)l}{2N+1}\right)=0
\end{equation}
Thus we have shown that the system of eigenvalues and eigenvector
\eq{eq_EW_n_EV} indeed solves the eigenvalue problem
\eq{eq_eigenvalue_prob}.

Next we show that the eigenvalues are distinct, positive and
non-zero. The fact that the eigenvalues are positive and non-zero
follows directly by inspection of \eq{eq_EW_n_EV} and by noting that
all viscoelastic constants are positive. It remains to be shown that
there are no multiple eigenvalues. This can be seen after
reformulating the expression for the eigenvalues as:
\begin{equation}    \lambda_l=1+\frac{\kappa-\Gamma}{\Gamma+4\sin^2\frac{\pi l}{2(2N+1)}}
\end{equation}
Since $1\leq l\leq 2N$, it holds for the argument of the
$\sin$-function that $0<\frac{\pi l}{2(2N+1)}<\frac{\pi}{2}$. In this
interval, the $\sin$-function increases monotonically and is
single-valued. For this reason, the eigenvalues, $\lambda_l$, are also
single-valued. The eigenvalues increase monotonically with $l$ if
$\kappa <\Gamma$ and decrease monotonically for increasing $l$ if the
opposite inequality holds.

Next we show that the eigenvectors are orthogonal and their length
is given by $v_l=\sqrt{(2N+1)/2}$. Consider the matrix of normalized
eigenvectors, \emph{\textbf{U}}, defined in the main text
\eq{eq_def_U}. By means of this matrix, the statement to be shown can
be recapitulated as $\emph{\textbf{U}}^T\emph{\textbf{U}}=I
\Leftrightarrow (\emph{\textbf{U}} \emph{\textbf{U}})_{k,m}
=\delta_{k,m}$. The second relation follows since $\emph{\textbf{U}}$
is obviously symmetric. In the following we will evaluate the square
of the matrix $\emph{\textbf{U}}$ by components:
\begin{equation}
\label{eq_proof_for_UU1}
 \begin{array}{rl}
  \displaystyle(\emph{\textbf{U}}\emph{\textbf{U}})_{k,m}&\displaystyle =\sum_{j=1}^{2N}\emph{\textbf{U}}_{k,j}\emph{\textbf{U}}_{j,m}\\[0.5cm]
     &\displaystyle =\frac{2}{2N+1}\sum_{j=1}^{2N}\sin\frac{\pi k j}{2N+1}\sin\frac{\pi j m}{2N+1}\\[0.5cm]
     &\displaystyle =\displaystyle\frac{1}{2N+1}\sum_{j=1}^{2N}\left(\cos\frac{\pi j(k-m)}{2N+1}-\cos\frac{\pi j (k+m)}{2N+1}\right)\\[0.5cm]
 \end{array}
\end{equation}
The finite sums over the $\cos$-functions can be evaluated by expressing it in terms of exponential functions. The used identity is:
\begin{equation}
\label{eq_sum_over_cos}
 \begin{array}{rcl}
  \displaystyle\sum_{j=0}^n\cos(j\alpha)&\displaystyle=&\displaystyle\frac{1}{2}\left(1+\frac{\sin(\alpha(n+1/2))}{\sin(\alpha/2)}\right)\\[0.5cm]
 \end{array}
\end{equation}
Application of \eq{eq_sum_over_cos} in order to simplify \eq{eq_proof_for_UU1} finally yields: 
\begin{equation}
\label{eq_proof_for_UU1_final}
\begin{array}{rl}  \displaystyle(\emph{\textbf{U}}\emph{\textbf{U}})_{k,m}&\displaystyle
=\frac{1}{2(2N+1)}\left(\sin\left((k-m)\pi\right)\cot\frac{(k-m)\pi}{2(2N+1)}\right.\\[0.5cm]   &\displaystyle\left.\hspace{0.3cm}-\sin((k+m)\pi)\cot\frac{(k+m)\pi}{2(2N+1)}\right.\\[0.5cm]
&\displaystyle\left.\hspace{0.3cm}+\cos((k+m)\pi)-\cos((k-m)\pi)\right)
\end{array}
\end{equation}
There are two cases, namely $k=m$ and $k\neq m$, that have to be
considered.

First assume that $k=m$. In this case, the last two
terms in \eq{eq_proof_for_UU1_final} just cancel out each other. The
$\sin$-function in the second term evaluates to zero while the
cotangent gives a finite value: since
$0<\frac{(k+m)\pi}{2(2N+1)}<\pi$, the singularities are just
spared. Thus, also this term vanishes. It is only the first term that
gives a contribution. Using l'H{\^{o}}pital's
rule, it evaluates to:
\begin{equation}
\label{eq_UU_diag}
\begin{array}{rl}  \displaystyle (\emph{\textbf{U}}\emph{\textbf{U}})_{k,k}&\displaystyle =\frac{1}{2(2N+1)}\lim_{m\rightarrow
k}\frac{\sin\left((k-m)\pi\right)}{\sin\frac{(k-m)\pi}{2(2N+1)}}\\[0.5cm]
& \displaystyle =\lim_{m\rightarrow k}\frac{\cos\left((k-m) \pi\right)} {\cos\frac{(k-m)\pi}{2(2N+1)}}=1
\end{array}
\end{equation}
This result ensures that all diagonal components of
$\emph{\textbf{U}}^2$ are unity.

Next assume that $k\neq m$. In
this case the first as well as the second term in
\eq{eq_proof_for_UU1_final} vanish since the $\sin$-function
evaluates to zero while the co-tangent yields finite values. The
last two terms further simplify to:
\begin{equation}
\begin{array}{rcl}
\label{eq_UU_offdiag}
  \displaystyle(\emph{\textbf{U}}\emph{\textbf{U}})_{k,m\neq k}&\displaystyle=&\displaystyle\frac{1}{2(2N+1)}\left((-1)^{k+m}-(-1)^{k-m}\right)\\[0.5cm]
   &\displaystyle=&\displaystyle\frac{1}{2(2N+1)}(-1)^{k-m}\left((-1)^{2m}-1)\right)=0
\end{array}
\end{equation}
This result ensures that all off-diagonal components of
$\emph{\textbf{U}}^2$ vanish. The combination of \eq{eq_UU_diag}
and \eq{eq_UU_offdiag} yields
$(\emph{\textbf{U}}\emph{\textbf{U}})_{k,m}=\delta_{k,m}$ which
was to be demonstrated. Thus we have shown that all
eigenvectors are of length $v_l=\sqrt{(2N+1)/2}$ and  form a
complete orthogonal basis.


\begin{thebibliography}{37}
\expandafter\ifx\csname natexlab\endcsname\relax\def\natexlab#1{#1}\fi
\expandafter\ifx\csname bibnamefont\endcsname\relax
  \def\bibnamefont#1{#1}\fi
\expandafter\ifx\csname bibfnamefont\endcsname\relax
  \def\bibfnamefont#1{#1}\fi
\expandafter\ifx\csname citenamefont\endcsname\relax
  \def\citenamefont#1{#1}\fi
\expandafter\ifx\csname url\endcsname\relax
  \def\url#1{\texttt{#1}}\fi
\expandafter\ifx\csname urlprefix\endcsname\relax\def\urlprefix{URL }\fi
\providecommand{\bibinfo}[2]{#2}
\providecommand{\eprint}[2][]{\url{#2}}

\bibitem[{\citenamefont{Gittes et~al.}(1997)\citenamefont{Gittes, Schnurr,
  Olmsted, MacKintosh, and Schmidt}}]{Gittes97}
\bibinfo{author}{\bibfnamefont{F.}~\bibnamefont{Gittes}},
  \bibinfo{author}{\bibfnamefont{B.}~\bibnamefont{Schnurr}},
  \bibinfo{author}{\bibfnamefont{P.~D.} \bibnamefont{Olmsted}},
  \bibinfo{author}{\bibfnamefont{F.~C.} \bibnamefont{MacKintosh}},
  \bibnamefont{and} \bibinfo{author}{\bibfnamefont{C.~F.}
  \bibnamefont{Schmidt}}, \bibinfo{journal}{Phys. Rev. Lett.}
  \textbf{\bibinfo{volume}{79}}, \bibinfo{pages}{3286} (\bibinfo{year}{1997}).

\bibitem[{\citenamefont{Gardel et~al.}(2004)\citenamefont{Gardel, Shin,
  MacKintosh, Mahadevan, Matsudaira, and Weitz}}]{Gardel04}
\bibinfo{author}{\bibfnamefont{M.~L.} \bibnamefont{Gardel}},
  \bibinfo{author}{\bibfnamefont{J.~H.} \bibnamefont{Shin}},
  \bibinfo{author}{\bibfnamefont{F.~C.} \bibnamefont{MacKintosh}},
  \bibinfo{author}{\bibfnamefont{L.}~\bibnamefont{Mahadevan}},
  \bibinfo{author}{\bibfnamefont{P.}~\bibnamefont{Matsudaira}},
  \bibnamefont{and} \bibinfo{author}{\bibfnamefont{D.~A.} \bibnamefont{Weitz}},
  \bibinfo{journal}{Science} \textbf{\bibinfo{volume}{304}},
  \bibinfo{pages}{1301} (\bibinfo{year}{2004}).

\bibitem[{\citenamefont{Mizuno et~al.}(2007)\citenamefont{Mizuno, Tardin,
  Schmidt, and MacKintosh}}]{Mizuno07}
\bibinfo{author}{\bibfnamefont{D.}~\bibnamefont{Mizuno}},
  \bibinfo{author}{\bibfnamefont{C.}~\bibnamefont{Tardin}},
  \bibinfo{author}{\bibfnamefont{C.~F.} \bibnamefont{Schmidt}},
  \bibnamefont{and} \bibinfo{author}{\bibfnamefont{F.~C.}
  \bibnamefont{MacKintosh}}, \bibinfo{journal}{Science}
  \textbf{\bibinfo{volume}{315}}, \bibinfo{pages}{370} (\bibinfo{year}{2007}).

\bibitem[{\citenamefont{Koenderink et~al.}(2009)\citenamefont{Koenderink,
  Dogic, Nakamura, Bendix, {MacKintosh}, Hartwig, Stossel, and
  Weitz}}]{koenderink_active_2009}
\bibinfo{author}{\bibfnamefont{G.~H.} \bibnamefont{Koenderink}},
  \bibinfo{author}{\bibfnamefont{Z.}~\bibnamefont{Dogic}},
  \bibinfo{author}{\bibfnamefont{F.}~\bibnamefont{Nakamura}},
  \bibinfo{author}{\bibfnamefont{P.~M.} \bibnamefont{Bendix}},
  \bibinfo{author}{\bibfnamefont{F.~C.} \bibnamefont{{MacKintosh}}},
  \bibinfo{author}{\bibfnamefont{J.~H.} \bibnamefont{Hartwig}},
  \bibinfo{author}{\bibfnamefont{T.~P.} \bibnamefont{Stossel}},
  \bibnamefont{and} \bibinfo{author}{\bibfnamefont{D.~A.} \bibnamefont{Weitz}},
  \bibinfo{journal}{Proc. Nat. Acad. Sci.}
  \textbf{\bibinfo{volume}{106}}, \bibinfo{pages}{15192 }
  (\bibinfo{year}{2009}).

\bibitem[{\citenamefont{Fabry et~al.}(2001)\citenamefont{Fabry, Maksym, Butler,
  Glogauer, Navajas, and Fredberg}}]{Fabry01}
\bibinfo{author}{\bibfnamefont{B.}~\bibnamefont{Fabry}},
  \bibinfo{author}{\bibfnamefont{G.~N.} \bibnamefont{Maksym}},
  \bibinfo{author}{\bibfnamefont{J.~P.} \bibnamefont{Butler}},
  \bibinfo{author}{\bibfnamefont{M.}~\bibnamefont{Glogauer}},
  \bibinfo{author}{\bibfnamefont{D.}~\bibnamefont{Navajas}}, \bibnamefont{and}
  \bibinfo{author}{\bibfnamefont{J.~J.} \bibnamefont{Fredberg}},
  \bibinfo{journal}{Phys. Rev. Lett.} \textbf{\bibinfo{volume}{87}},
  \bibinfo{pages}{148102} (\bibinfo{year}{2001}).

\bibitem[{\citenamefont{Fern{\'a}ndez et~al.}(2006)\citenamefont{Fern{\'a}ndez,
  Pullarkat, and Ott}}]{Fernandez06}
\bibinfo{author}{\bibfnamefont{P.}~\bibnamefont{Fern{\'a}ndez}},
  \bibinfo{author}{\bibfnamefont{P.~A.} \bibnamefont{Pullarkat}},
  \bibnamefont{and} \bibinfo{author}{\bibfnamefont{A.}~\bibnamefont{Ott}},
  \bibinfo{journal}{Biophys. J.} \textbf{\bibinfo{volume}{90}},
  \bibinfo{pages}{3796} (\bibinfo{year}{2006}).

\bibitem[{\citenamefont{Discher et~al.}(2005)\citenamefont{Discher, Janmey, and
  Wang}}]{Discher05}
\bibinfo{author}{\bibfnamefont{D.~E.} \bibnamefont{Discher}},
  \bibinfo{author}{\bibfnamefont{P.}~\bibnamefont{Janmey}}, \bibnamefont{and}
  \bibinfo{author}{\bibfnamefont{Y.-L.} \bibnamefont{Wang}},
  \bibinfo{journal}{Science} \textbf{\bibinfo{volume}{310}},
  \bibinfo{pages}{1139} (\bibinfo{year}{2005}).

\bibitem[{\citenamefont{Vogel and Sheetz}(2006)}]{vogel_local_2006}
\bibinfo{author}{\bibfnamefont{V.}~\bibnamefont{Vogel}} \bibnamefont{and}
  \bibinfo{author}{\bibfnamefont{M.}~\bibnamefont{Sheetz}},
  \bibinfo{journal}{Nat. Rev. Mol. Cell Biol.} \textbf{\bibinfo{volume}{7}},
  \bibinfo{pages}{265} (\bibinfo{year}{2006}).

\bibitem[{\citenamefont{Geiger et~al.}(2009)\citenamefont{Geiger, Spatz, and
  Bershadsky}}]{geiger_environmental_2009}
\bibinfo{author}{\bibfnamefont{B.}~\bibnamefont{Geiger}},
  \bibinfo{author}{\bibfnamefont{J.~P.} \bibnamefont{Spatz}}, \bibnamefont{and}
  \bibinfo{author}{\bibfnamefont{A.~D.} \bibnamefont{Bershadsky}},
  \bibinfo{journal}{Nat. Rev. Mol. Cell Biol.} \textbf{\bibinfo{volume}{10}},
  \bibinfo{pages}{21} (\bibinfo{year}{2009}).

\bibitem[{\citenamefont{Pellegrin and Mellor}(2007)}]{pellegrin_actin_2007}
\bibinfo{author}{\bibfnamefont{S.}~\bibnamefont{Pellegrin}} \bibnamefont{and}
  \bibinfo{author}{\bibfnamefont{H.}~\bibnamefont{Mellor}}, \bibinfo{journal}{J.
  Cell Sci.} \textbf{\bibinfo{volume}{120}}, \bibinfo{pages}{3491}
  (\bibinfo{year}{2007}).

\bibitem[{\citenamefont{Peterson et~al.}(2004)\citenamefont{Peterson, Rajfur,
  Maddox, Freel, Chen, Edlund, Otey, and Burridge}}]{Peterson04}
\bibinfo{author}{\bibfnamefont{L.~J.} \bibnamefont{Peterson}},
  \bibinfo{author}{\bibfnamefont{Z.}~\bibnamefont{Rajfur}},
  \bibinfo{author}{\bibfnamefont{A.~S.} \bibnamefont{Maddox}},
  \bibinfo{author}{\bibfnamefont{C.~D.} \bibnamefont{Freel}},
  \bibinfo{author}{\bibfnamefont{Y.}~\bibnamefont{Chen}},
  \bibinfo{author}{\bibfnamefont{M.}~\bibnamefont{Edlund}},
  \bibinfo{author}{\bibfnamefont{C.}~\bibnamefont{Otey}}, \bibnamefont{and}
  \bibinfo{author}{\bibfnamefont{K.}~\bibnamefont{Burridge}},
  \bibinfo{journal}{Mol. Biol. Cell}
  \textbf{\bibinfo{volume}{15}}, \bibinfo{pages}{3497} (\bibinfo{year}{2004}).

\bibitem[{\citenamefont{Endlich et~al.}(2007)\citenamefont{Endlich, Otey, Kriz,
  and Endlich}}]{endlich_movement_2007}
\bibinfo{author}{\bibfnamefont{N.}~\bibnamefont{Endlich}},
  \bibinfo{author}{\bibfnamefont{C.}~\bibnamefont{Otey}},
  \bibinfo{author}{\bibfnamefont{W.}~\bibnamefont{Kriz}}, \bibnamefont{and}
  \bibinfo{author}{\bibfnamefont{K.}~\bibnamefont{Endlich}},
  \bibinfo{journal}{Cell Mot. Cytoskel.}
  \textbf{\bibinfo{volume}{64}}, \bibinfo{pages}{966} (\bibinfo{year}{2007}).

\bibitem[{\citenamefont{Smith et~al.}(2010)\citenamefont{Smith, Blankman,
  Gardel, Luettjohann, Waterman, and Beckerle}}]{smith_zyxin-mediated_2010}
\bibinfo{author}{\bibfnamefont{M.}~\bibnamefont{Smith}},
  \bibinfo{author}{\bibfnamefont{E.}~\bibnamefont{Blankman}},
  \bibinfo{author}{\bibfnamefont{M.}~\bibnamefont{Gardel}},
  \bibinfo{author}{\bibfnamefont{L.}~\bibnamefont{Luettjohann}},
  \bibinfo{author}{\bibfnamefont{C.}~\bibnamefont{Waterman}}, \bibnamefont{and}
  \bibinfo{author}{\bibfnamefont{M.}~\bibnamefont{Beckerle}},
  \bibinfo{journal}{Dev. Cell} \textbf{\bibinfo{volume}{19}},
  \bibinfo{pages}{365} (\bibinfo{year}{2010}).

\bibitem[{\citenamefont{Katoh et~al.}(1998)\citenamefont{Katoh, Kano, Masuda,
  Onishi, and Fujiwara}}]{katoh_isolation_1998}
\bibinfo{author}{\bibfnamefont{K.}~\bibnamefont{Katoh}},
  \bibinfo{author}{\bibfnamefont{Y.}~\bibnamefont{Kano}},
  \bibinfo{author}{\bibfnamefont{M.}~\bibnamefont{Masuda}},
  \bibinfo{author}{\bibfnamefont{H.}~\bibnamefont{Onishi}}, \bibnamefont{and}
  \bibinfo{author}{\bibfnamefont{K.}~\bibnamefont{Fujiwara}},
  \bibinfo{journal}{Mol. Biol. Cell}
  \textbf{\bibinfo{volume}{9}}, \bibinfo{pages}{1919} (\bibinfo{year}{1998}).

\bibitem[{\citenamefont{Deguchi et~al.}(2006)\citenamefont{Deguchi, Ohashi, and
  Sato}}]{Deguchi06}
\bibinfo{author}{\bibfnamefont{S.}~\bibnamefont{Deguchi}},
  \bibinfo{author}{\bibfnamefont{T.}~\bibnamefont{Ohashi}}, \bibnamefont{and}
  \bibinfo{author}{\bibfnamefont{M.}~\bibnamefont{Sato}}, \bibinfo{journal}{J.
  Biomech.} \textbf{\bibinfo{volume}{39}}, \bibinfo{pages}{2603}
  (\bibinfo{year}{2006}).

\bibitem[{\citenamefont{Matsui et~al.}(2009)\citenamefont{Matsui, Deguchi,
  Sakamoto, Ohashi, and Sato}}]{Matsui09}
\bibinfo{author}{\bibfnamefont{T.}~\bibnamefont{Matsui}},
  \bibinfo{author}{\bibfnamefont{S.}~\bibnamefont{Deguchi}},
  \bibinfo{author}{\bibfnamefont{N.}~\bibnamefont{Sakamoto}},
  \bibinfo{author}{\bibfnamefont{T.}~\bibnamefont{Ohashi}}, \bibnamefont{and}
  \bibinfo{author}{\bibfnamefont{M.}~\bibnamefont{Sato}},
  \bibinfo{journal}{Biorheology} \textbf{\bibinfo{volume}{46}},
  \bibinfo{pages}{401} (\bibinfo{year}{2009}).

\bibitem[{\citenamefont{Kumar et~al.}(2006)\citenamefont{Kumar, Maxwell,
  Heisterkamp, Polte, Lele, Salanga, Mazur, and
  Ingber}}]{kumar_viscoelastic_2006}
\bibinfo{author}{\bibfnamefont{S.}~\bibnamefont{Kumar}},
  \bibinfo{author}{\bibfnamefont{I.}~\bibnamefont{Maxwell}},
  \bibinfo{author}{\bibfnamefont{A.}~\bibnamefont{Heisterkamp}},
  \bibinfo{author}{\bibfnamefont{T.}~\bibnamefont{Polte}},
  \bibinfo{author}{\bibfnamefont{T.}~\bibnamefont{Lele}},
  \bibinfo{author}{\bibfnamefont{M.}~\bibnamefont{Salanga}},
  \bibinfo{author}{\bibfnamefont{E.}~\bibnamefont{Mazur}}, \bibnamefont{and}
  \bibinfo{author}{\bibfnamefont{D.}~\bibnamefont{Ingber}},
  \bibinfo{journal}{Biophys. J.} \textbf{\bibinfo{volume}{90}},
  \bibinfo{pages}{3762} (\bibinfo{year}{2006}).

\bibitem[{\citenamefont{Colombelli et~al.}(2009)\citenamefont{Colombelli,
  Besser, Kress, Reynaud, Girard, Caussinus, Haselmann, Small, Schwarz, and
  Stelzer}}]{uss:colo09}
\bibinfo{author}{\bibfnamefont{J.}~\bibnamefont{Colombelli}},
  \bibinfo{author}{\bibfnamefont{A.}~\bibnamefont{Besser}},
  \bibinfo{author}{\bibfnamefont{H.}~\bibnamefont{Kress}},
  \bibinfo{author}{\bibfnamefont{E.}~\bibnamefont{Reynaud}},
  \bibinfo{author}{\bibfnamefont{P.}~\bibnamefont{Girard}},
  \bibinfo{author}{\bibfnamefont{E.}~\bibnamefont{Caussinus}},
  \bibinfo{author}{\bibfnamefont{U.}~\bibnamefont{Haselmann}},
  \bibinfo{author}{\bibfnamefont{J.}~\bibnamefont{Small}},
  \bibinfo{author}{\bibfnamefont{U.~S.} \bibnamefont{Schwarz}},
  \bibnamefont{and} \bibinfo{author}{\bibfnamefont{E.}~\bibnamefont{Stelzer}},
  \bibinfo{journal}{J. Cell Sci.} \textbf{\bibinfo{volume}{122}},
  \bibinfo{pages}{1665} (\bibinfo{year}{2009}).

\bibitem[{\citenamefont{Tanner et~al.}(2010)\citenamefont{Tanner, Boudreau,
  Bissell, and Kumar}}]{tanner_dissecting_2010}
\bibinfo{author}{\bibfnamefont{K.}~\bibnamefont{Tanner}},
  \bibinfo{author}{\bibfnamefont{A.}~\bibnamefont{Boudreau}},
  \bibinfo{author}{\bibfnamefont{M.}~\bibnamefont{Bissell}}, \bibnamefont{and}
  \bibinfo{author}{\bibfnamefont{S.}~\bibnamefont{Kumar}},
  \bibinfo{journal}{Biophys. J.} \textbf{\bibinfo{volume}{99}},
  \bibinfo{pages}{2775} (\bibinfo{year}{2010}).

\bibitem[{\citenamefont{Kruse and Julicher}(2000)}]{kruse_actively_2000}
\bibinfo{author}{\bibfnamefont{K.}~\bibnamefont{Kruse}} \bibnamefont{and}
  \bibinfo{author}{\bibfnamefont{F.}~\bibnamefont{Julicher}},
  \bibinfo{journal}{Phys. Rev. Lett.}
  \textbf{\bibinfo{volume}{85}}, \bibinfo{pages}{1778} (\bibinfo{year}{2000}).

\bibitem[{\citenamefont{Kruse and
  Julicher}(2003)}]{kruse_self-organization_2003}
\bibinfo{author}{\bibfnamefont{K.}~\bibnamefont{Kruse}} \bibnamefont{and}
  \bibinfo{author}{\bibfnamefont{F.}~\bibnamefont{Julicher}},
  \bibinfo{journal}{Phys. Rev. E} \textbf{\bibinfo{volume}{67}}
  (\bibinfo{year}{2003}).

\bibitem[{\citenamefont{Besser and Schwarz}(2007)}]{Besser07}
\bibinfo{author}{\bibfnamefont{A.}~\bibnamefont{Besser}} \bibnamefont{and}
  \bibinfo{author}{\bibfnamefont{U.~S.} \bibnamefont{Schwarz}},
  \bibinfo{journal}{New J. Phys.} \textbf{\bibinfo{volume}{9}},
  \bibinfo{pages}{425} (\bibinfo{year}{2007}).

\bibitem[{\citenamefont{Stachowiak and O'Shaughnessy}(2008)}]{Stachowiak08}
\bibinfo{author}{\bibfnamefont{M.~R.} \bibnamefont{Stachowiak}}
  \bibnamefont{and}
  \bibinfo{author}{\bibfnamefont{B.}~\bibnamefont{O'Shaughnessy}},
  \bibinfo{journal}{New J. Phys.} \textbf{\bibinfo{volume}{10}},
  \bibinfo{pages}{025002} (\bibinfo{year}{2008}).

\bibitem[{\citenamefont{Luo et~al.}(2008)\citenamefont{Luo, Xu, Lele, Kumar,
  and Ingber}}]{Luo08}
\bibinfo{author}{\bibfnamefont{Y.}~\bibnamefont{Luo}},
  \bibinfo{author}{\bibfnamefont{X.}~\bibnamefont{Xu}},
  \bibinfo{author}{\bibfnamefont{T.}~\bibnamefont{Lele}},
  \bibinfo{author}{\bibfnamefont{S.}~\bibnamefont{Kumar}}, \bibnamefont{and}
  \bibinfo{author}{\bibfnamefont{D.~E.} \bibnamefont{Ingber}},
  \bibinfo{journal}{J. Biomech.} \textbf{\bibinfo{volume}{41}},
  \bibinfo{pages}{2379} (\bibinfo{year}{2008}).

\bibitem[{\citenamefont{Stachowiak and O'Shaughnessy}(2009)}]{Stachowiak09}
\bibinfo{author}{\bibfnamefont{M.~R.} \bibnamefont{Stachowiak}}
  \bibnamefont{and}
  \bibinfo{author}{\bibfnamefont{B.}~\bibnamefont{O'Shaughnessy}},
  \bibinfo{journal}{Biophys. J.} \textbf{\bibinfo{volume}{97}},
  \bibinfo{pages}{462} (\bibinfo{year}{2009}).

\bibitem[{\citenamefont{Russell et~al.}(2009)\citenamefont{Russell, Xia,
  Dickinson, and Lele}}]{Russell09}
\bibinfo{author}{\bibfnamefont{R.}~\bibnamefont{Russell}},
  \bibinfo{author}{\bibfnamefont{S.}~\bibnamefont{Xia}},
  \bibinfo{author}{\bibfnamefont{R.}~\bibnamefont{Dickinson}},
  \bibnamefont{and} \bibinfo{author}{\bibfnamefont{T.}~\bibnamefont{Lele}},
  \bibinfo{journal}{Biophys. J.} \textbf{\bibinfo{volume}{97}},
  \bibinfo{pages}{1578} (\bibinfo{year}{2009}).

\bibitem[{\citenamefont{Besser and Schwarz}(2010)}]{uss:bess10a}
\bibinfo{author}{\bibfnamefont{A.}~\bibnamefont{Besser}} \bibnamefont{and}
  \bibinfo{author}{\bibfnamefont{U.~S.} \bibnamefont{Schwarz}},
  \bibinfo{journal}{Biophys. J.} \textbf{\bibinfo{volume}{99}},
  \bibinfo{pages}{L10} (\bibinfo{year}{2010}).

\bibitem[{\citenamefont{Pipkin}(1986)}]{b:pipkin86}
\bibinfo{author}{\bibfnamefont{A.}~\bibnamefont{Pipkin}},
  \emph{\bibinfo{title}{Lectures on Viscoelasticity Theory}}, Applied
  Mathematical Sciences (\bibinfo{publisher}{Springer}, \bibinfo{address}{New
  York}, \bibinfo{year}{1986}).

\bibitem[{\citenamefont{Howard}(2001)}]{Howard01}
\bibinfo{author}{\bibfnamefont{J.}~\bibnamefont{Howard}},
  \emph{\bibinfo{title}{{Mechanics of motor proteins and the cytoskeleton}}}
  (\bibinfo{publisher}{Sunderland, Sinauer Associates}, \bibinfo{year}{2001}).

\bibitem[{\citenamefont{Kargin and Slonimsky}(1948)}]{Kargin48}
\bibinfo{author}{\bibfnamefont{V.~A.} \bibnamefont{Kargin}} \bibnamefont{and}
  \bibinfo{author}{\bibfnamefont{G.~L.} \bibnamefont{Slonimsky}},
  \bibinfo{journal}{Doklady Akademii Nauk SSSR} \textbf{\bibinfo{volume}{62}},
  \bibinfo{pages}{239} (\bibinfo{year}{1948}).

\bibitem[{\citenamefont{Kargin and Slonimsky}(1949)}]{Kargin49}
\bibinfo{author}{\bibfnamefont{V.~A.} \bibnamefont{Kargin}} \bibnamefont{and}
  \bibinfo{author}{\bibfnamefont{G.~L.} \bibnamefont{Slonimsky}},
  \bibinfo{journal}{Zurnal Fiziceskoj Chimii} \textbf{\bibinfo{volume}{23}},
  \bibinfo{pages}{563} (\bibinfo{year}{1949}).

\bibitem[{\citenamefont{Rouse}(1953)}]{Rouse53}
\bibinfo{author}{\bibfnamefont{P.~E.} \bibnamefont{Rouse}},
  \bibinfo{journal}{The Journal of Chemical Physics}
  \textbf{\bibinfo{volume}{21}}, \bibinfo{pages}{1272} (\bibinfo{year}{1953}).

\bibitem[{\citenamefont{Vinogradov and Malkin}(1980)}]{VinogradovMalkin80}
\bibinfo{author}{\bibfnamefont{G.~V.} \bibnamefont{Vinogradov}}
  \bibnamefont{and} \bibinfo{author}{\bibfnamefont{A.~Y.}
  \bibnamefont{Malkin}}, \emph{\bibinfo{title}{Rheology of polymers:
  Viscoelasticity and flow of polymers}} (\bibinfo{publisher}{Springer-Verlag},
  \bibinfo{address}{Berlin}, \bibinfo{year}{1980}).

\bibitem[{\citenamefont{Gotlib and Volkenshtein}(1953)}]{Gotlib53}
\bibinfo{author}{\bibfnamefont{Y.~Y.} \bibnamefont{Gotlib}} \bibnamefont{and}
  \bibinfo{author}{\bibfnamefont{M.~V.} \bibnamefont{Volkenshtein}},
  \bibinfo{journal}{Zurnal Techniceskoj Fiziki} \textbf{\bibinfo{volume}{23}},
  \bibinfo{pages}{1936} (\bibinfo{year}{1953}).

\bibitem[{\citenamefont{Coughlin and Stamenovic}(2003)}]{c:coug03}
\bibinfo{author}{\bibfnamefont{M.~F.} \bibnamefont{Coughlin}} \bibnamefont{and}
  \bibinfo{author}{\bibfnamefont{D.}~\bibnamefont{Stamenovic}},
  \bibinfo{journal}{Biophys J.} \textbf{\bibinfo{volume}{84}},
  \bibinfo{pages}{1328} (\bibinfo{year}{2003}).

\bibitem[{\citenamefont{Paul et~al.}(2008)\citenamefont{Paul, Heil, Spatz, and
  Schwarz}}]{uss:paul08a}
\bibinfo{author}{\bibfnamefont{R.}~\bibnamefont{Paul}},
  \bibinfo{author}{\bibfnamefont{P.}~\bibnamefont{Heil}},
  \bibinfo{author}{\bibfnamefont{J.~P.} \bibnamefont{Spatz}}, \bibnamefont{and}
  \bibinfo{author}{\bibfnamefont{U.~S.} \bibnamefont{Schwarz}},
  \bibinfo{journal}{Biophys. J.} \textbf{\bibinfo{volume}{94}},
  \bibinfo{pages}{1470} (\bibinfo{year}{2008}).

\bibitem[{\citenamefont{Bischofs et~al.}(2008)\citenamefont{Bischofs, Klein,
  Lehnert, Bastmeyer, and Schwarz}}]{uss:bisc08a}
\bibinfo{author}{\bibfnamefont{I.~B.} \bibnamefont{Bischofs}},
  \bibinfo{author}{\bibfnamefont{F.}~\bibnamefont{Klein}},
  \bibinfo{author}{\bibfnamefont{D.}~\bibnamefont{Lehnert}},
  \bibinfo{author}{\bibfnamefont{M.}~\bibnamefont{Bastmeyer}},
  \bibnamefont{and} \bibinfo{author}{\bibfnamefont{U.~S.}
  \bibnamefont{Schwarz}}, \bibinfo{journal}{Biophys. J.}
  \textbf{\bibinfo{volume}{95}}, \bibinfo{pages}{3488–3496}
  (\bibinfo{year}{2008}).

\end{thebibliography}

\end{document}